\documentstyle[11pt,epsf,rotate]{article}
\newcommand{\bce}{\begin{center}}
\newcommand{\ece}{\end{center}}
\newcommand{\be}{\begin{equation}}
\newcommand{\ee}{\end{equation}}
\newcommand{\bea}{\begin{eqnarray}}
\newcommand{\eea}{\end{eqnarray}}
\newcommand{\bit}{\begin{itemize}}
\newcommand{\eit}{\end{itemize}}
\newcommand{\E}{\>=\>}
\newcommand{\EA}{&=&}
\newcommand{\To}{\> \longrightarrow \> }
\newcommand{\TO}{\> \stackrel{T \to \infty}{\longrightarrow} \> }
\newcommand{\bfl}{\begin{flushright}}
\newcommand{\efl}{\end{flushright}}

\renewcommand{\thesection}{\arabic{section}}

\renewcommand{\theequation}{\thesection.\arabic{equation}}

\def\PRstyle#1#2#3#4#5{{#1} {\bf #2#3} (#4) #5}

\def\NUCPHYS{ Nucl. Phys.}
\def\PHYSLETT{ Phys. Lett.}
\def\PHYSREP{ Phys. Rep.}
\def\PRL{ Phys. Rev. Lett.}
\def\PR{ Phys. Rev.}
\def\ZEITPHYS{ Z. Phys.}

\def\IJMP{ Int. J. Mod. Phys.}
\def\ANNPHYS{ Ann. Phys.}
\def\JOURPHYS{ J. Phys.}

\begin{document}
\thispagestyle{empty}

\vspace{2cm}
%

\bce
{\large \bf Improved Variational Description of the
Wick-Cutkosky Model with the Most General Quadratic Trial 
Action}
\vspace{2cm}

\vspace{1.2cm}
R. Rosenfelder $^1$ and A. W. Schreiber $^2$

\vspace{0.8cm}
$^1$ Paul Scherrer Institute, CH-5232 Villigen PSI, Switzerland

\vspace{0.1cm}
$^2$ Department of Physics and Mathematical Physics, and
           Research Centre for the Subatomic Structure of Matter,
           University of Adelaide, Adelaide, S. A. 5005, Australia

\ece
%
\vspace{2.5cm}

\begin{abstract}
\noindent

We generalize the worldline variational approach to field theory by
introducing a trial action which allows for anisotropic terms to be
induced by external 4-momenta of Green's functions.  By solving the
ensuing variational equations numerically we demonstrate that within
the (quenched) scalar Wick-Cutkosky model considerable improvement can
be achieved over results obtained previously with isotropic actions.
In particular, the critical coupling associated with the instability
of the model is lowered, in accordance with expectations from Baym's
proof of the instability in the unquenched theory. The physical
picture associated with a different quantum mechanical motion of
the dressed particle along and perpendicular to its classical momentum
is discussed.  Indeed, we find that for large couplings the
dressed particle is strongly distorted in the direction of its
four-momentum.  In addition, we obtain an exact relation between
the renormalized coupling of the theory and the propagator.  Along the way
we
introduce new and efficient methods to
evaluate the averages needed in the variational approach and apply
them to the calculation of the 2-point function.
\end{abstract}

\newpage
\section{Introduction}
\label{sec: introduction}
\setcounter{equation}{0}

\vspace{1cm}

\noindent
Feynman's variational method for evaluating functional integrals, well
known in condensed matter physics~\cite{Feyn}, makes use of a trial
action containing variational parameters. Even though the exact action
of the theory under consideration can in general be quite complicated,
the trial action itself can be at most quadratic in the degrees of
freedom of the theory because the only functional integrals which can
be performed analytically are Gaussian ones.  Clearly this imposes a
significant constraint so that one might at first sight suspect that
such a variational method, even if valid in principle, is likely to be
of limited use in practice.  This turns out not to be the case,
however: Feynman formulated the variational calculation on the level
of a non-local action for the polaron, obtained after integrating out
the phonon degrees of freedom.  This non-locality, which is shared by
the trial action used by him, introduces considerable freedom for the
variational principle to work with and hence he obtained numerical
results for, e.g. the polaron's groundstate energy, which differed --
for a very large range of coupling constants -- from the exact ones by
at most 2 \% percent.

Encouraged by this success we have extended this method to relativistic 
field theory ~\cite{WC1} -- \cite{WC6} following the pioneering, but 
largely forgotten, work by K. Mano ~\cite{Mano}. In these investigations 
Feynman's approach was applied to a
relativistic scalar field theory of two fields $\varphi$ and $\Phi$
with Lagrangian (in Minkowski space)
\be
{\cal L} \E \frac{1}{2} \left ( \partial_{\mu} \Phi \right )^2 -
\frac{1}{2}
M_0^2 \Phi^2 + \frac{1}{2} \left ( \partial_{\mu} \varphi \right )^2
 - \frac{1}{2}m^2 \varphi^2 + g \, \Phi^2 \varphi 
\label{def WC}
\ee
where we refer to the light field $\varphi$ as ``meson'' and to the 
heavy one ($\Phi$) as ``nucleon''. 
This theory is the Wick-Cutkosky model \cite{Wick,Cut} and is of 
interest
as it models, in a rather rough fashion, a simplified nucleon-pion
theory without the complications of spin and isospin (and chiral
symmetry). Due to the close similarity to the polaron model (indeed,
the dressed nucleon might be referred to as ``relativistic polaron'', 
the only
essential difference being the space-time dimensionality) one would
expect the variational method to work equally well in this setting.
Despite of this similarity, there are noticable qualitative differences:
most importantly, while the polaron is a stable (quasi-) particle at all
couplings,  it turns out (see Refs.~\cite{WC1,WC2}) that
the variational equations of the Wick-Cutkosky model
only have (real) solutions below a
critical coupling of $\alpha_{c} \approx 0.815$, where
\be
\alpha \E  {g^2 \over 4 \pi M^2}
\label{eq: def alpha}
\ee
is the dimensionless coupling constant and $M$ the physical mass
associated with the field $\Phi$.  The existence of a critical
coupling is likely to be a remnant of the well-known instability of
cubic scalar field theories~\cite{Baym}.

\vspace{0.5cm}
There has been renewed interest in this theory and its instability in
recent years.  Ahlig and Alkofer~\cite{Alkofer} have analysed the
stability of the model using Dyson-Schwinger equation methods and have
found a critical coupling rather similar to the one mentioned above.
Tjon and coworkers~\cite{Tjon1} have studied the theory (without the 
radiative
corrections which are responsible for the instability) using
Monte-Carlo techniques and have recently also alluded to evidence of a 
critical coupling~\cite{Tjon2}.  It has also been suggested by these
authors~\cite{Tjon3} that the apparent stability of the quenched
theory below some critical coupling is a real effect 
\footnote{In our opinion, however, this suggestion cannot be
 correct as it conflicts with standard expectations based on the
 behaviour of high orders of the perturbative expansion of the theory:
 The contribution from the $K^{th}$ order to the Euclidean coordinate
 space 2-point function grows roughly like the number of Feynman
 diagrams at that order.  This growth is factorial in both the
 unquenched and quenched theory.  Moreover, as the expansion is in
 $g^2$ rather than $g$, the contribution from each order is of the same
 sign.  Therefore the series is not Borel summable, indicating that a
 perturbative expansion is an expansion around the wrong vacuum.
 Quantitative calculations summarized in Appendix~\ref{app: lopt}
 confirm this expectation. Also, it seems that the ``proof'' 
 in~\cite{Tjon3}
 is flawed {\it inter alia} by wrongly asserting that the true mass is 
 {\it larger} than the variationally estimated value which invalidates 
 all subsequent steps.}
and not just a byproduct of approximations~\cite{comment}.  
In principle, the ``true'' $\alpha_c$ in the quenched theory could be 
found numerically by Monte-Carlo simulation similar to
the way the true ground state energy of the polaron has been determined
by stochastic methods \cite{MC polaron}. 
Notwithstanding the interest in this critical coupling, we also note that 
there has been a great deal of interest in
recent years in field theories with imaginary or negative coupling 
 constants
(which would lead to stability even for the unquenched theory) which,
although they have non-hermitian Hamiltonians, are nevertheless argued
to have a positive-definite spectrum by virtue of their 
PT-symmetry~\cite{Bender}.
Finally, of course, there is continued interest in using this field 
theory as a vehicle for examining boundstate problems~\cite{boundstate}.

In recent applications we have treated a more realistic 
fermionic theory, {\it viz.} Quantum Electrodynamics, by worldline 
variational methods \cite{QED2} and obtained a compact
non-perturbative expression
for the anomalous mass dimension. This shows that this approach 
successfully 
describes the short-distance singularities
of relativistic quantum field theories. It is remarkable that at the
same time the long-distance behaviour is also caught to a large extent 
as evidenced by the proper threshold behaviour of scattering amplitudes 
in the scalar model \cite{WC3,WC5} or the correct exponentiation of 
infrared singularities in QED \cite{ARS}. 

Encouraging as these specific results are, one would like to assess the 
reliability of the variational results in general.
There are two
obvious ways one can do this: the approximation at the heart of
Feynman's variational calculation is the cumulant expansion and
corrections can be systematically calculated, as was done for the
polaron in Ref. ~\cite{polaron corr}.  Alternatively, one can make use
of more general trial actions, thus allowing the variational principle
more freedom to work with.  It is the latter course of action which we
pursue here, concentrating for simplicity on the simple scalar field 
theory
defined in Eq. (\ref{def WC}) and considered in  Refs.~\cite{WC1,WC2}. 
The results presented there employed
a number of different forms for the trial action, culminating in the
most general (isotropic), non-local, quadratic trial action possible.
It might seem, because the trial action 
is required to be quadratic, that it is hard to improve on this.
However, as has been explored within the context of the polaron in
Ref.~\cite{polaron_aniso}, one can make use of the external direction
provided by the particle's four-momentum to construct more general,
{\it anisotropic}, trial actions.  For the polaron this extra freedom only
yields marginally improved results due to the nonrelativistic nature
of that problem~\cite{polaron_aniso}.  One might expect however, and
we shall indeed find, that significant improvements can be achieved in
the present, relativistic, theory.

The organization of this paper is as follows.  In the next Section, 
after briefly summarizing the essential details of the variational
method (for more details, we refer the reader to Ref.~\cite{WC1}), we
describe the anisotropic trial action and derive the relevant
variational equations.  In Sec.~\ref{sec: numerics} we present
numerical results and in Sec.~\ref{sec: conclusion} we conclude. 
Technical details are collected in five appendices.

\section{The variational method}
\label{sec: formalism}
\setcounter{equation}{0}

\subsection{Feynman's variational method applied to the Wick-Cutkosky
model and the anisotropic trial action}

In the polaron variational approach the key elements are

\bit

\item[i)]{The reduction of degrees of freedom by using the particle 
(``worldline'')
representation and by integrating out the mesons/photons. For example, 
in the 
one-nucleon sector of the scalar theory defined by Eq. (\ref{def WC}) 
this is achieved by  
using the Feynman-Schwinger representation (as well as neglecting the
functional determinant 
if one is working in the quenched approximation)
\be
\frac{1}{-\partial^2 - M_0^2 + 2 g \varphi(x)+ i 0} \E - \frac{i}{2 
\kappa_0}
\int_0^{\infty} dT \> 
\exp \left [ \frac{i}{2 \kappa_0} \left (-\partial^2 - M_0^2 + 
2 g \varphi(x) \right ) \, T \, \right ] \> .
\label{FS repr}
\ee
Here $\kappa_0$ is a free (positive) parameter which reparametrizes the 
proper time 
$T$ without affecting the physics \footnote{It also allows an easy 
(formal) switch to 
Euclidean space by setting  $\kappa_0 = i \kappa_E $ and changing the 
sign of 4-vector products. Proper times remain unchanged and the action 
$S$ becomes $iS_E$.}. 
The exponential
factor on the r.h.s. of Eq. (\ref{FS repr}) can be considered as a 
proper-time
evolution operator and may be represented as a quantum mechanical path 
integral. 
With $\varphi(x)$ now being a c-number quantity the integration
over the mesonic field can be performed resulting in an effective action 
for the four-dimensional trajectory $x(t)$ of the nucleon
\bea
S[x] \EA \int_{t_0}^{t_0+T} dt \left ( - \frac{\kappa_0}{2} \right ) 
\dot x^2(t)  
- \frac{g^2}{2 \kappa_0^2} \int_{t_0}^{t_0+T} dt_1\int_{t_0}^{t_0+T} dt_2
\, \int \frac{d^4 k}{(2 \pi)^4} \> \frac{1}{k^2 - m^2 + i0} \nonumber \\
&& \hspace{5cm} \cdot \exp \left \{ - i k \cdot \left [ x(t_1) - x(t_2) 
\right ] \, \right \} \> \equiv \> S_0 + S_1 \> .
\label{S0+S1}
\eea
Here the starting time $t_0$ is free since only the proper time 
interval $T$ matters. Convenient choices are $t_0 = 0$ or $t_0 = -T/2$.
}

\item[ii)]{The use of the Feynman-Jensen variational principle
\bea
\int {\cal D} x \, e^{i S_t } \, \exp \left [ \, i \left ( S  - S_t 
\right ) 
\, \right ] &\equiv&   \left (\int {\cal D} x \, e^{i S_t } \right )
\, \cdot \, \langle  \, \exp \left [i \left ( S - S_t \right ) \right ] \,
\rangle_{S_t} \nonumber\\
&\simeq&   \left ( \int {\cal D} x \, e^{i S_t }\right ) \, \cdot \,
 \exp \left [ \, i \langle \,  S - S_t  \, \rangle_{S_t} \, \right ] \, ,
\label{FeynJens}
\eea
where $<\ldots>_{S_t}$ refers to averaging with weight function
$e^{i S_t }$ (normalized to $<1>_{S_t}=1$) and $ \> \simeq \> $ indicates 
equality at the stationary point of 
the r. h. s. under unrestricted variations of the trial action 
$S_t$. }

\item[iii)]{A quadratic two-time trial action 
\be
S_t^{\rm iso \> \> (1)} \E  S_0 + \frac{i \kappa_0^2}{2} 
\int_{t_0}^{t_0+T} dt_1dt_2 \> f (t_1-t_2) \> \Bigl [ \> x(t_1) -
 x(t_2) \> \Bigr ]^2\;\;\;,
\label{eq: iso St 1}
\ee
in which the non-quadratic terms in the true action $S$ are approximated 
by an even {\it retardation function} $f(\sigma)$. Since $t_0$ is 
arbitrary 
(proper time-translation invariance) this function can only depend on the 
time difference $\sigma = t_1- t_2$. Note that the true action as well 
as the
trial action are invariant under a constant translation in $x$. Therefore 
one also could take
\be
S_t^{\rm iso \> \> (2)} \E  S_0 - i \kappa_0^2 
\int_{t_0}^{t_0+T} dt_1dt_2 \> g (t_1-t_2) \> \dot x(t_1) 
\cdot \dot x(t_2) \> ,
\label{eq: iso St 2}
\ee
because two integration by parts and neglect of boundary terms show that 
this is equivalent to Eq. (\ref{eq: iso St 1}) with $f(\sigma) = 
\ddot g(\sigma)$.
}

\eit
Being mainly interested in the on-shell-limit of the Green functions of 
the theory
it is convenient to include the Fourier transform over the endpoints 
into the 
path integral (``momentum averaging'' in the parlance of 
Ref. ~\cite{WC1}) and to consider
\be 
\tilde S[x] \E p \cdot x \> + S[x] \> .
\label{eq: tilde S}
\ee 
The nucleon's propagator may be calculated from the generating functional 
and the worldline representation via
\be
G_2(p) \E \frac{1}{2 i \kappa_0} \> \int_0^{\infty} dT \>
\exp \left [ \frac{i T}{2 \kappa_0} ( p^2 - M_0^2)  \right ] \> \cdot \>
\underbrace{
{\int {\cal D}{\tilde x} \> e^{i \tilde S[x]} \over \int 
{\cal D}{\tilde x} 
\>e^{i \tilde S_0[x]} } }_{=: g_2(p,T)}\;\;\;.
\label{eq: propagator}
\ee
Note that we have normalized the functional measure  by dividing by
the path integral involving
the free action $S_0$ (i.e. Eq.~(\ref{eq: tilde S}) with $g=0$).
As mentioned,     the path integrals in Eq.~(\ref{eq: propagator}) 
include an integral over the endpoint $x$ -- hence the tilde over the $x$.

\noindent
The variational approximation to this propagator, making use of
Jensen's stationary principle, has the form (\ref{eq: propagator}) but 
with
\be
g_2(p,T) \> \simeq \> g_2^{\rm var}(p,T) \E 
{\int {\cal D}{\tilde x}\> e^{i \tilde S_t[x]} \over \int 
{\cal D}{\tilde x} \>e^{i \tilde S_0[x]} } \, 
\exp\left [\, i\,
{\int {\cal D}{\tilde x}\> ( \tilde S[x]- \tilde S_t[x]) \; e^{i 
\tilde S_t[x]} 
\over \int {\cal D}{\tilde x}\> e^{i \tilde S_t[x]} } \, \right] \;\;.
\label{eq: g2 var}
\ee 
Of course, when the extended trial action $\tilde S_t$ equals
$\tilde S$ this expression gives the exact propagator in 
Eq.~(\ref{eq: propagator}).  
On the other hand, as
long as $\tilde S_t$ is quadratic in $x(t)$, all functional integrals
appearing in Eq.~(\ref{eq: g2 var}) may be performed
analytically. The most general linear + quadratic isotropic trial action, 
respecting symmetries such as translational invariance, may be
written as 
\be
\tilde S_t^{iso}[x] \E \tilde \lambda\, p \cdot x + S_t^{\rm iso} 
\label{eq: tilde St}
\ee
where $\tilde \lambda$ is an additional variational parameter.
Note that the structure of the linear term is basically fixed by
requiring that the action is a scalar and by time translation invariance:
there is only one additional four-vector, the external momentum $p$, with 
which the velocity four-vector can be contracted and any modification of
the free term $ p \cdot x = p \cdot \int dt \, v(t) $ can only be done by 
multiplication with a constant, but not with a function of the proper 
time. 
In contrast, the retardation functions $f$ or $g$ in the quadratic 
term must be functions of the time difference $ \sigma = t_1- t_2$.

The key to the success of Feynman's variational method is that the
different functional dependence on the path $x(t)$ in the exact action
in Eq.~(\ref{eq: tilde S}) as compared to the trial action in
Eq.~(\ref{eq: iso St 1}) can be compensated for by the
variational retardation function $f(\sigma)$.  For
example, in the path integral in Eq.~(\ref{eq: propagator}) contributions
from $[x(t_1)-x(t_2)]^2 \approx 0$ are greatly enhanced because of
the UV divergence in the interacting part of the action 
Eq.~(\ref{eq: tilde S}).  In the path integral in 
Eq.~(\ref{eq: g2 var}) the
equivalent enhancement is provided by a divergence of $f(\sigma)$ as
$\sigma \rightarrow 0$.  Similarly, infrared physics is simulated
by the large-$\sigma$ behaviour of this function.  However, 
in general it is to be expected that in the functional integral
over $\, \exp ( i S[x] ) \, $ paths which contain segments 
$x_\mu(t_1)-x_\mu(t_2)$ 
which are parallel
to the momentum $p_\mu$ will receive a different weighting
to those containing segments which are perpendicular to $p_\mu$.
Because $f(\sigma)$ is scalar, the trial actions 
~(\ref{eq: iso St 1}, \ref{eq: iso St 2}) 
cannot differentiate between these two.  Rather, the
behaviour of the
retardation function $f(\sigma)$ will encapsulate some compromise
of this directional information. 

The situation is different, however, if we generalize the trial
action by explicitly making use of the external vector $p_\mu$.
The most general covariant,
quadratic, anisotropic trial action may be written as
\bea
S_{t}^{\rm aniso \> 1} \EA  S_0 + \frac{i \kappa_0^2}{2} \,
  \int_{t_0}^{t_0+T} dt_1dt_2 \;\left[\, f_L(\sigma) 
{p^\mu p^\nu \over p^2} + f_T(\sigma)
\left ( g^{\mu \nu} - {p^\mu p^\nu \over p^2}\right) \, \right ] 
\nonumber \\
&&\hspace{4.cm} \cdot \biggl[ \, x_\mu(t_1) - x_\mu(t_2) \biggr ] \>  
\biggl[ x_\nu(t_1) - x_\nu(t_2) \, \biggr ]\;\;\;.
\label{eq: aniso St 1}
\eea
In this action we now have two independent retardation functions
$f_T(\sigma)$  and $f_L(\sigma)$ and we shall show, in the following,
that these indeed encode  just
that directional information in the functional integral which we have 
discussed above. Equivalently, just as in the isotropic case 
(i.e. Eq.~(\ref{eq: iso St 2})), one could take
\be
S_{t}^{\rm aniso \> 2} \E  S_0 - i \kappa_0^2 \, 
  \int_{t_0}^{t_0+T} dt_1dt_2 \;\left[ \, g_L(\sigma) {p^\mu p^\nu 
\over p^2} + g_T(\sigma)
\left ( g^{\mu \nu} - {p^\mu p^\nu \over p^2}\right) \, \right ] \, 
\dot x_\mu(t_1) \, \dot x_\nu(t_2) \> 
\label{eq: aniso St 2}\\
\ee
instead. 

\subsection{Mano's equation and functional averages}

In Appendix {\ref{app: average} it is detailed how 
the averages required in Eq.~(\ref{eq: g2 var}) 
may be calculated easily by a method which is more transparent and 
efficient 
than the Fourier series expansion used in Ref.~\cite{WC1}. In particular, 
Eq. (\ref{tilde St}) demonstrates that all trial actions may be written 
as 
\be 
\tilde S_t[v] \E - \frac{\kappa_0}{2} \, (v | A | v) + (b | v )
\ee
if expressed in terms of {\it velocities} $v$. Allowing for a Lorentz 
structure of $A$,
i.e. $(v | A | v) \equiv (v_\mu | A^{\mu \nu} | v_\nu)$,
this is obviously the most general linear + quadratic trial action. 
In principle
one could work out the required averages with this trial action and let 
the variational principle determine the form of $A$ and $b$. 
That this indeed leads to an anisotropic trial action is 
sketched in Appendix \ref{app: finite T}.
However, since there is only one
four-vector available for the 2-point function, namely the external 
momentum $p$, 
it is clear that $b$ must be proportional to $p$ and that $A$ must have 
the 
decomposition (\ref{A par perp}), which reflects the Lorentz structure 
of the 
retardation functions in Eqs. (\ref{eq: aniso St 1}, 
\ref{eq: aniso St 2}).

In the following we will only consider the limit of the 2-point function 
where $p^2 \to M^2$, i.e. the on-mass-shell limit.
It is well known \cite{trunc} that in order to obtain the pole and 
the residue of the nucleon propagator
\be
G_2(p) \To \frac{Z}{p^2 - M^2 + i 0} \> , 
\label{G2 at pole}
\ee
the proper time $T$ has to tend to infinity. To obtain the physical mass 
in terms of the bare mass only the leading large-$T$ term in the 
exponential 
of $g_2$ is of relevance (see Eq. (\ref{log g2 large T})
of Appendix \ref{app: large T}). Performing the
$T$-integration in Eq. (\ref{eq: propagator}) and setting $p^2 = M^2$ 
gives
\be
M_0^2 \E (2 \lambda - \lambda^2) M^2 - 2 (\Omega + V) \> .
\label{eq: Mano}
\ee
This equation, whose content is discussed in detail below, was termed 
`Mano's equation' in Ref.~\cite{WC1}.
There it was also shown that for fixed
physical mass $M$ (and fixed UV regulator) the variational principle  
in fact provides a lower bound on the exact bare mass, i.e.
\be
M_0^2({\rm exact})\> \ge \>M_0^2 \, \left ({\rm Eq.}~(\ref{eq: Mano}) 
\right )\;\;\;.
\ee
The parameter $\lambda$ in Mano's equation (\ref{eq: Mano}) is related 
to the 
original variational parameter $\tilde \lambda$ by
\be
\lambda \E \frac{\tilde \lambda}{A_L(0)} \> .
\label{eq: def lambda}
\ee
As shown in Eq. (\ref{Omega aniso})
those averages not involving the coupling $\alpha$ are contained in the 
`kinetic term' $\Omega$, which in $d$ dimensions is given by
\be
\Omega[A_T,A_L] \E \frac{1}{d} \Omega[A_L] + {d-1\over d} \, \Omega[A_T] 
  \;\;\;,
\ee
with
\be
\Omega[A_{L,T}] \E  {d \kappa_0 \over 2 i \pi} \int_0^\infty dE \> \left [
\log A_{L,T}(E) \> + \> {1\over A_{L,T}(E)} \> - \> 1\right ]\;\;\;.
\ee
Note that when $A_T=A_L$ this reduces to the isotropic result and that the
factor $d-1$  in front of the first term arises because in $d$ dimensions 
there
are $d-1$ dimensions transverse to $p_\mu$.  From Eqs. (\ref{A(E) from 
f(sigma)}, \ref{A(E) from g(sigma)}) it is seen that
the transverse and longitudinal profile functions
$A_{L,T}$ are essentially the cosine transforms of the 
transverse and longitudinal retardation functions, respectively.
Since $\log A + 1/A -1$ is positive for $A > 0$ the kinetic term provides 
the restoring force to balance the attractive interaction.

The `potential' term, essentially the average of the interaction dependent
part of the action, is given by Eq. (\ref{V aniso}) which in $d=4$ 
dimensions reads
\bea
V \EA -\frac{g^2}{8\pi^2} \, \int_0^{\infty} d \sigma \, \int_0^1du \> 
\Biggl \{ {F_{LT}^{3/2}(u,\sigma) \over \mu_L^2(\sigma)}
\> \exp \left [ -{i \over 2 \kappa_0 } \left ( m^2 \mu_L^2(\sigma) 
{1-u \over u} +
{\lambda^2 M^2 \sigma^2 \over \mu_L^2(\sigma) }u \right ) \right ] 
\nonumber \\
&&\hspace{4cm}  - \> {1\over \sigma} \>
\exp \left[ -{i \over 2\kappa_0} m^2 \sigma {1-u \over
u} \right]  \Biggr \}
\> - \> {g^2 \over 8 \pi^2} \log {\Lambda^2 \over m^2}.
\label{eq: v}
\eea
In addition, we have used a Pauli-Villars regulator as in Ref.~\cite{WC1} to 
make the proper time integral convergent for small $\sigma$.
As in Refs.~\cite{WC1,WC2} the divergent logarithm on the right 
hand side will be taken over to the
left hand side of Eq.~(\ref{eq: Mano}), with the replacement of $M_0^2$
by the finite 
\be
M_1^2 \E M_0^2 - {g^2 \over 4 \pi^2} \log {\Lambda^2 \over m^2} \> .
\label{def M1}
\ee
The essential new element in Eq.~(\ref{eq: v}) as compared to
the isotropic case is the appearance of the anisotropy factor
\be
F_{LT}(u,\sigma) \E \left [ \, 1+ \left ( {\mu_T^2(\sigma) \over
\mu_L^2(\sigma)} - 1\right ) u \, \right ]^{-1} \;\;\;.
\ee
Also, note that there is now one `pseudotime' $\mu^2(\sigma)$ for
each of $A_{L,T}(E)$:
\be
\mu^2_{L,T}(\sigma) \> = \> {4 \over \pi} \int_0^\infty {dE \over E^2}
\> {\sin^2{E \sigma \over 2} \over A_{L,T}(E)}\;\;\;.
\label{eq: pseudotimes}
\ee
This function was named as such in Ref.~\cite{WC1} because, as can be
seen rather straightforwardly from Eq.~(\ref{eq: pseudotimes}), it is
proportional to $\sigma$ ($\sigma > 0$) for small $\sigma$ as well as for
large $\sigma$ (and, indeed, is equal to $\sigma$ when the interaction
is turned off, i.e. when $A(E)=1$).  Actually,
these pseudotimes also have a direct physical interpretation: Let
us consider, initially, the average separation $x(t_1)-x(t_2)$
when weighted with the exponential of the trial action.  Using Eqs. 
(\ref{exp average},  \ref{a1 large T})
this is easily calculated and one obtains
\be
\langle \, x^{\mu}(t_1)-x^{\mu}(t_2)  \, \rangle_{ \tilde S_t} \E - 
\frac{a_1^{\mu}}{\kappa_0} \> 
\TO \> { \lambda p^\mu  \over \kappa_0}\;\sigma\;\;\;.
\ee
The meaning of this expression is clear:  the
average (four-) displacement that the nucleon undertakes in the proper 
time
interval $\sigma=t_1-t_2$ is given by its (four-) velocity $\cdot$ 
$\sigma$.
It is natural then to view $\kappa_0/\lambda$ as an effective mass 
\footnote{
Eq. (\ref{S0+S1}) shows that $\kappa_0$ is the ``mass'' of the particle 
in the 
worldline description. Integration over the proper time $T$ with the
weight $\exp(-i M_0^2T/(2 \kappa_0))$ gives the particle its actual 
(bare) mass.
Note that in the nonrelativistic limit the proper time
can be identified with the ordinary time when $\kappa_0 = M$ is chosen 
\cite{QED1}.}.
That this
interpretation is not at all unreasonable was seen for the 3-dimensional
polaron problem discussed in Ref.~\cite{polaron_aniso}, where the variational 
equations for $\lambda$ directly lead to this result.

\noindent
It is very instructive to consider now the {\em mean square} displacement.
With Eq. (\ref{a2 large T}) one obtains
\be
\langle  \, \left[ x^{\mu}(t_1)- x^{\mu}(t_2) \right] \, 
\left[ x^{\nu}(t_1)-x^{\nu}(t_2) \right] \, \rangle_{ S_t} 
\E \frac{1}{\kappa_0^2} a_1^{\mu} a_1^{\nu} - \frac{i}{\kappa_0} a_2^{\mu \nu}
\> \TO \>   \frac{\lambda^2 p^{\mu} p^{\nu}}{\kappa_0^2}  \sigma^2
 - \frac{i}{\kappa_0} \left ( \mu^2(\sigma) \right )^{\mu \nu} \> .
\label{eq: mean square disp}
\ee

\unitlength1mm
\begin{figure}[ht]
\begin{center}
\mbox{\epsfysize=6cm\epsffile{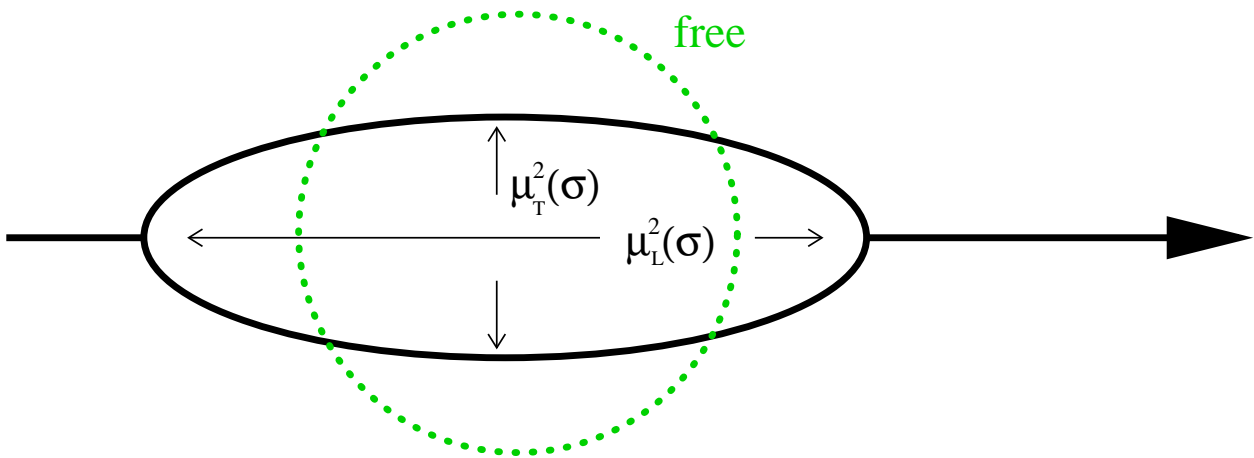}}
\end{center}

\caption{The physical interpretation of
the pseudotimes $\mu_{L,T}(\sigma)$.  The dotted 
circle denotes the noninteracting
particle with quantum mechanical uncertainty in $ \langle x^2 \rangle $ given
by Brownian motion alone.}
\label{fig: pseudotimes}
\end{figure}

The first term is again the contribution from the straight line
(i.e. classical) motion.  The last term characterizes both the quantum
mechanical deviations from this as well as the `jiggling' the nucleon
experiences because of the constant emission and re-absorption of the
pions.  For example, as mentioned above the pseudotimes reduce to
$\sigma$ and $\lambda = 1$ when the interactions are turned off.  In
this case the first term in Eq.~(\ref{eq: mean square disp})
represents a constant drift and the second just corresponds to the
result well known from Brownian motion that the mean square distance
(in each direction) grows linearly with Euclidean time (see footnote
2).  Turning on the interactions modifies this Brownian motion, and
 since 
\be
\left ( \mu^2(\sigma) \right )^{\mu \nu} 
\E \mu^2_L(\sigma) \, \frac{p^{\mu} p^{\nu}}{p^2} 
+ \mu^2_T(\sigma) \left (  g^{\mu \nu}
- \frac{p^{\mu} p^{\nu}}{p^2} \right ) 
\ee
this modification may be different in the longitudinal and transverse
directions.

The results are summarized in the cartoon in Fig.~\ref{fig:
pseudotimes}, anticipating the numerical results from Section
\ref{sec: numerics}.  Note that the possibility of distortion in the
longitudinal direction is only present because the trial action has
been allowed to be anisotropic. It is shown in Appendix \ref{app:
finite T} that for nonvanishing momentum $p$ the variational
principle, if given sufficient freedom, {\it demands} such a
structure.

\subsection{The variational equations}

By demanding that the independent variation of Mano's equation  
(ie. Eq.~(\ref{eq: Mano})) with respect to the parameter 
$\lambda$ and the profile
functions $A_T(E)$ and $A_L(E)$ must vanish, we obtain the
variational equations for these functions. Since a numerical solution
is only feasible in Euclidean space we choose $\kappa_0 = i \kappa_E$. 
After some work one obtains
\be
{1 \over \lambda} \E 1 \> + \>{g^2 \over 8 \pi^2} \, \frac{1}{\kappa_E} 
\int_0^\infty d \sigma {\sigma^2 \over \mu_L^4(\sigma)}
\int_0^1du \> u \> F_{LT}^{3/2}(u,\sigma) \> \cdot  e(u,\sigma)
\label{eq: var lambda}
\ee

\bea
A_T(E) \EA 1 + \frac{8}{3 \kappa_E E^2} \,\int_0^\infty d\sigma \> 
\frac{\delta V}{\delta \mu^2_T(\sigma)} \, \sin^2 {E \sigma \over 2} 
\nonumber \\
\frac{\delta V}{\delta \mu^2_T(\sigma)} \EA 
{3g^2 \over 16 \pi^2} \, \frac{1}{\mu_L^4(\sigma)}
\int_0^1 du \> u \, F_{LT}^{5/2}(u,\sigma) \> \cdot
e(u,\sigma)
\label{eq: var AT}
\eea

\bea
A_L(E) \EA 1 + \frac{8}{\kappa_E E^2} \, \int_0^\infty d\sigma \> 
\frac{\delta V}{\delta \mu^2_L(\sigma)} \, \sin^2 {E \sigma \over 2} 
\nonumber\\
\frac{\delta V}{\delta \mu^2_L(\sigma)} \EA {g^2 \over 16 \pi^2}
\frac{1}{\mu_L^4(\sigma)}
\int_0^1 du \> u \, F_{LT}^{3/2}(u,\sigma) \, 
\left( 1 - \frac{1}{\kappa_E} 
{\lambda^2 M^2 \sigma^2 \over \mu_L^2(\sigma) }u \right) \> \cdot
e(u,\sigma) 
\label{eq: var AL}
\eea
where
\be
e(u,\sigma) \> \equiv \> \exp \left [ \, - \frac{1}{2 \kappa_E}
\left ( m^2 \mu_L^2(\sigma) \frac{1-u}{u} + \frac{\lambda^2 M^2 
\sigma^2}{\mu_L^2(\sigma)} u \right ) \, \right ] \> .
\ee
Note that the compact form of the last variational equation results 
from an appropriate integration by parts in $u$. Neither Eq. 
(\ref{eq: var AT}) nor
Eq. (\ref{eq: var AL}) reduce to the variational equation for the isotropic 
profile function given in Sec. IV. C of Ref. \cite{WC1} for $F_{LT} = 1$ 
because
the anisotropy factor $F_{LT}$ has been varied as well. However, 
both equations
become identical when $M=0$ and $F_{LT}=1$.  This just corresponds
to the fact that when the momentum $p_\mu$ vanishes 
(hence $p^2=-M^2=0$) there is no source of anisotropy and hence
no need for separate retardation functions in the longitudinal
and transverse directions. In this case one also deduces from Eqs. 
(\ref{eq: var AL})
and (\ref{eq: var lambda}) that $\lambda A(0) = 1 $, i.e. the original 
variational
parameter $\tilde \lambda$ in the trial action (\ref{eq: tilde St}) decouples 
and is unaffected by the interaction.

\noindent
Finally, by comparison with Eq. (\ref{A(E) from f(sigma)}) 
one also sees that the retardation functions defined in 
Eq.~(\ref{eq: aniso St 1})
are essentially the variational derivatives of the `potential'
\be
f^{\rm var}_{L,T} (\sigma) \E \frac{c_{L,T}}{\kappa_E^2} \, 
\frac{\delta V}{\delta \mu^2_{L,T}(\sigma)} \> , \> \> \> \> c_L \E 1 \> , 
\> \> 
\> \> c_T \E \frac{1}{3} \> .
\ee

\subsection{The residue}
For completeness we also evaluate the residue of the propagator in 
Eq. (\ref{G2 at pole}). This requires the calculation of the next-to-leading 
terms in the large-$T$ limit. As shown in Appendix \ref{app: T NLO} only the 
interaction gives a ${\cal O}(T^0)$-contribution to $ \> \log g_2(p,T) \> $. 
Therefore following
Sec. V. A in Ref. \cite{WC2} and using Eq. (\ref{V next}) we easily obtain 
in $d = 4$ Euclidean dimensions 
\be
Z \E \frac{1}{\lambda} \,  \exp \left [ \, -  \frac{g^2}{8 \pi^2} \, 
\frac{1}{\kappa_E}\,\int_0^{\infty} d\sigma
\, \frac{\sigma}{\mu_L^2(\sigma)} \int_0^1 du \> F_{LT}^{3/2} (u,\sigma) 
\cdot e(u,\sigma) \, \right ] \> \equiv \> \frac{N}{\lambda} \> .
\label{eq: residue}
\ee
As in  Ref. \cite{WC2}, the variational equation (\ref{eq: var lambda}) for 
$\lambda$ has been used to simplify the denominator of this expression.
It is also easily seen that $Z$ does not depend on the value of $\kappa_E$.

\noindent
Apart from the obvious modifications due to the anisotropic trial
action and the general reparametrization gauge $\kappa_E \neq 1$,
Eq. (\ref{eq: residue}) differs from the result obtained in
Ref. \cite{WC2} by the absence of the factor $ N_0 = \exp[ -\log A(0)
+ 1 - 1/A(0) ] $. As seen in Eq. (65) of Ref. \cite{WC2} this factor
had its origin in a ${\cal O}(1/T)$-correction to the kinetic term
$\Omega$ which does not occur in the present formulation where an
even, time-translation invariant retardation function (or
equivalently, an even profile function $A(t-t')$) has been employed
throughout. The discrepancy is explained by recalling that -- for
practical reasons -- the previous trial action was taken as a
quadratic, {\it diagonal} form in Fourier space (see Eqs. (48) - (51)
in Ref. \cite{WC1}). After transforming back to $x$-space it may be
seen that this form also contains terms which are not invariant under
time translations.  The present formulation therefore not only
respects this obvious symmetry of the action but also eliminates the
spurious and awkward $N_0$-term \footnote{It should be noted that all
previous results for physical amplitudes are independent of $N_0$, as
can be seen, e.g. in Eq. (52) of Ref. \cite{WC4}.}: after absorbing
$A(0)$ in the variational parameter $\lambda$, it does not appear
explicitly anymore, neither in the pole mass nor in the residue of the
propagator.

\subsection{The effective coupling}

Finally, although it is beyond the scope of this paper to discuss
the vertex function in general (for the isotropic trial actions this 
was done in
Ref.~\cite{WC4}), it is rather straightforward to derive the
effective coupling of the theory (i.e. the value of the vertex function 
at $q^2=0$) for an {\it arbitrary} trial action. 

We start with the (exact) worldline representation of the untruncated
3-point function, which is closely related to the corresponding
expression for the propagator, i.e. Eq.~(\ref{eq: propagator}).  The
only difference is an additional plane wave $e^{i q\cdot x(\tau)}$ for
the external pion with (outgoing) momentum $q^\mu$ and an integral
over the proper time $\tau$ at which the pion couples to the nucleon's
worldline:
\begin{equation}
G_{2,1}(p,p'=p-q) \, =  \, {\rm const.}\>  \int_0^\infty dT
\exp \left [ - {i\,T \over 2\kappa_0} M_0^2 \right ]
 \int {\cal D} \tilde x
\> e^{- i p'\cdot x}
\left [ g \int_{0}^{T} d\tau \>
e^{i q \cdot x(\tau)}\right ] e^{i \tilde S[x]}\;\;.
\label{eq: g21}
\end{equation}
For $q^\mu=0$ the integral over the time $\tau$ just provides a
factor of $T$, while the rest of the integrand is the same
as the one for the propagator in Eq.~(\ref{eq: propagator}).  We
therefore obtain
\be
G_{2,1}(p,p)\> = \> 2\, g \,{\partial \over \partial M_0^2}\> G_2(p)\;\;,
\label{eq: g21_q0}
\ee
where the normalization has been fixed so that the correct
free limit is obtained (see Ref.~\cite{WC4}; the factor
$2$ arises because of the definition of the coupling
in Eq.~(\ref{def WC}) without a factor $1/2!$) and we have made
use of the fact that the only dependence on the bare mass
enters through the explicit exponential factor shown in 
Eq.~(\ref{eq: g21}).  
One merely has to truncate the external legs off $G_{2,1}(p,p')$
and multiply by the residue $Z$ in order to obtain the vertex function.
At $q_{\mu}=0$ this defines the effective coupling $2\, g_{\rm phys}$.
Hence the physical coupling of the theory and the propagator are
related by the (exact) relation
\be
g_{\rm phys} \> = \> -\, g\, Z \,
{\partial \over \partial M_0^2}\> G_2^{-1}(p)\;\;\;.
\label{eq: eff coup}
\ee
In order to obtain the variational result for this,
we write the propagator near the mass shell as
\be
G_2(p) \> = \> {N \over {\cal M}(M^2) + (p^2-M^2) \lambda}\;\;\;.
\label{eq: propdef}
\ee
Here $N$ is defined in Eq.~(\ref{eq: residue}) and
${\cal M}(M^2) = (2 \lambda - \lambda^2) M^2 - 2 (\Omega + V) - M_0^2 = 0 $
is Mano's equation (\ref{eq: Mano})
\footnote{We need to momentarily retain ${\cal M}(M^2)$ because of its 
dependence on the bare mass.}. Note that the denominator is just the first few
terms of the Taylor expansion of ${\cal M}(p^2)$ around
$p^2 = M^2$.  The equality ${\cal M}'(p^2)|_{p^2=M^2} = \lambda$ 
only makes use
of the variational equation for $\lambda$ and hence is independent
of whether an isotropic or anisotropic trial action is used.
Substituting into Eq.~(\ref{eq: eff coup}) immediately
yields the variational estimate for the effective coupling
\be
g_{\rm phys} \> = \> {g \over \lambda}\;\;\;.
\ee
As we know from previous calculations and also shall see in the next section, 
$\lambda$ is below $1$ and
hence the physical coupling is larger than the bare coupling.

\section{Numerical Results}
\label{sec: numerics}
\setcounter{equation}{0}

We have solved the variational equations (\ref{eq: var
lambda})~--~(\ref{eq: var AL}) numerically for a variety of
dimensionless couplings $\alpha$ (defined in Eq. (\ref{eq: def alpha})) 
while keeping
the physical mass of the nucleon (pion) fixed at $M=939$ MeV ($m=140$
MeV). As in the previous work we haven chosen the Euclidean reparametrization
``gauge'' $\kappa_E = 1$ which only affects the $E, \sigma$-scale of profile 
functions and pseudotimes, whereas 
$\, \lambda, \, A_{L,T}(0), \, \Omega_{L,T}, \, V \, $
and therefore also the masses $M_0, \, M_1$ do not depend on
the reparametrization parameter. 
The numerical integrations were based on Gauss-Legendre
integration, after suitable mapping of the infinite integration ranges
to finite ones. The precision of the numerics was controlled by
demanding that results do not change appreciably upon subdivision of the
integration range.  The equations were solved by iteration, with the
convergence criterion being 1 part in $10^6$. In addition, as in the polaron 
case \cite{polaron_aniso}, we have used the {\it virial theorem} to
calculate the kinetic term in a completely different way:
\be
 \Omega^{\rm var} \> \equiv \> \frac{1}{4} \Omega^{\rm var}_L + 
\frac{3}{4} \Omega^{\rm var}_T  
\E  - \int_0^{\infty} d\sigma \, \sigma^2 \left [ \,
\frac{\delta V}{\delta \mu^2_L(\sigma)} \, \frac{\partial}{\partial \sigma}
\left (  \frac{\mu^2_L(\sigma)}{\sigma} \right ) 
+ \frac{\delta V}{\delta \mu^2_T(\sigma)} \,  \frac{\partial}{\partial \sigma}
\left (  \frac{\mu^2_T (\sigma)}{\sigma} \right ) \, \right ]\> .
\label{virial}
\ee
As shown in Appendix \ref{app: virial} this relation relies on the fact 
that profile functions and pseudotimes
are solutions of the variational equations and thereby represents a
crucial test for the accuracy of the numerical calculation. Unfortunately, in 
the present case the derivative of the pseudotimes cannot be eliminated but
has to be calculated numerically from
\be
\frac{\partial \mu^2_{L,T}}{\partial \sigma} \E 1 + \frac{2}{\pi} 
\int_0^{\infty} dE \> \frac{\sin(E \sigma)}{E} \, \left ( \, 
\frac{1}{A_{L,T}(E)} - 1 \, \right )
\ee
which -- due to less damping of the oscillating integrand than for 
the pseudotimes themselves -- gives less stable results than for 
$\mu^2_{T,L}$. 
Nevertheless, relative agreement better than  $ 3 \cdot 10^{-4} $ 
was achieved for $\Omega$ and better than
$ 6 \cdot 10^{-6} $ for $M_1$ in the whole range of coupling constants. 
Numerical results for $\lambda$ and the values of the profile
functions at $E=0$, as well as the resulting values for the mass $M_1$ and 
the residue $Z$, are tabulated in Table~\ref{tab: numerical results} for a
variety of the bare couplings $\alpha$.  Also tabulated are the corresponding 
values for the physical coupling $\alpha_{\rm phys}\>=\>\alpha/\lambda^2$.

\begin{table}[!htb]
\vspace{-5mm}
\begin{center}
{\bf Anisotropic Action}\\
\vspace{2mm}
\begin{tabular}{|c|cclllc|} 
\hline
     &          &           &         &        &                & \\
$\alpha$ & $\alpha_{\rm phys}$ &
$\lambda$ & $A_L(0)$  &  $A_T(0)$&  $M_1 $(MeV) &  Z \\
     &          &           &         &        &              & \\
\hline
     &          &           &         &        &                & \\
0.1  &  0.10566 &  0.97287 &   0.97688 &  1.02821 & 890.302 & 0.96051 \\
0.2  &  0.22472 &  0.94339 &   0.94996 &  1.06154 & 840.041 & 0.91749\\
0.3  &  0.36159 &  0.91086 &   0.91781 &  1.10207 & 788.106 & 0.86983 \\
0.4  &  0.52357 &  0.87406 &   0.87796 &  1.15355 & 734.416 & 0.81561 \\
0.5  &  0.72456 &  0.83071 &   0.82554 &  1.22369 & 679.01  & 0.75115 \\
0.6  &  0.99842 &  0.77521 &   0.74794 &  1.33366 & 622.28  & 0.66716\\
0.7  &  1.53567 &  0.67515 &   0.5675  &  1.6310  & 566.31  & 0.5075~~\\
     &          &           &         &        &               &  \\
\hline
\end{tabular}
\end{center}
\begin{center}
{\bf Isotropic Actions}\\
\vspace{2mm}
\begin{tabular}{|c|ccllc|cccc|} 
\hline
 &          &           &         &        &         &         &        &        
& \\
     & \multicolumn{5}{c|}{Best Variational}& \multicolumn{4}{c|}{Feynman}  \\
     &          &           &         &        &         &         &        &      
& \\
$\alpha$ & $\alpha_{\rm phys}$ &   $\lambda$ &
$A(0)$ & $M_1 $(MeV)& Z & $\alpha_{\rm phys}$ &   $\lambda$& $M_1 $(MeV)& Z  \\
     &          &           &         &        &         &         &       &        
& \\
\hline
     &          &           &         &        &         &         &       &        
& \\
0.1  &	0.10563 &   0.97297 & 1.01508 & 890.246 & 0.96087  & 0.10830 & 0.96090 & 
890.23 & 0.96090\\
0.2  &	0.22449 &   0.94389 & 1.03221 & 839.785 & 0.91918  & 0.23663 & 0.91934 & 
839.73 & 0.91934\\
0.3  &	0.36051 &   0.91223 & 1.05202 & 787.429 & 0.87428  & 0.39213 & 0.87467 & 
787.29 & 0.87467\\
0.4  &	0.51986 &   0.87718 & 1.07551 & 732.971 & 0.82521  & 0.58627 & 0.82600 & 
732.69 & 0.82600\\
0.5  &	0.71306 &   0.83738 & 1.10439 & 676.20  & 0.77036  & 0.83930 & 0.77184 & 
675.70 & 0.77184\\
0.6  &	0.96066 &   0.79030 & 1.14207 & 616.98 & 0.70672   & 1.1923  & 0.70940 & 
616.09 & 0.70940 \\
0.7  &	1.31472 &   0.72968 & 1.1972  & 555.48 & 0.62697   & 1.7516  & 0.63216 & 
553.93 & 0.63216\\
0.8  &	2.06369 &   0.62262 & 1.3188  & 493.55  & 0.49284  & 3.0654  & 0.51086 & 
490.60 & 0.51086\\
     &          &           &         &        &         &         &       &        
&  \\
\hline
\end{tabular}
\end{center}
\caption{{\it Top:} The variational parameter $\lambda$, the variational
profile functions at $E=0$, the modified bare mass $M_1$ and the residue
$Z$ obtained with the anisotropic trial
action.  Also shown is the value of the effective renormalized coupling
$\alpha_{\rm phys}$.
{\it Bottom:} For comparison,  
the corresponding results obtained
with the best isotropic action and with Feynman's parametrization 
of the retardation function are collected in the second Table
(see Tables III and I in Ref.~\protect\cite{WC2}). In all cases, 
the nucleon mass has been taken as $939$ MeV and the pion mass as $140$ MeV.}
\label{tab: numerical results}
\end{table}
\noindent
We draw the reader's attention to the fact
that for large $\sigma$ the pseudotimes $\mu^2_{L,T}(\sigma)$
behave like $\sigma / A_{L,T}(0)$.  
As is evident from the
Table, as the coupling constant is increased the 
behaviour for these pseudotimes is rather different 
in the longitudinal and transverse directions, leading
(for large $\sigma$) to a strong `elongation' 
of the dressed nucleon in the direction of (the 4-dimensional) motion and
a `contraction' in the perpendicular directions
(for the largest coupling, this distortion is depicted in 
Fig.~\ref{fig: pseudotimes}). Clearly,
the results obtained for $A(0)$ with the best isotropic actions (see the second
part of Table~\ref{tab: numerical results}) were a compromise 
between $A_T(0)$ and $A_L(0)$.

\unitlength1mm
\begin{figure}[!ht]
\vspace{-2cm}
\begin{center}
\mbox{\epsfxsize=14cm\epsffile{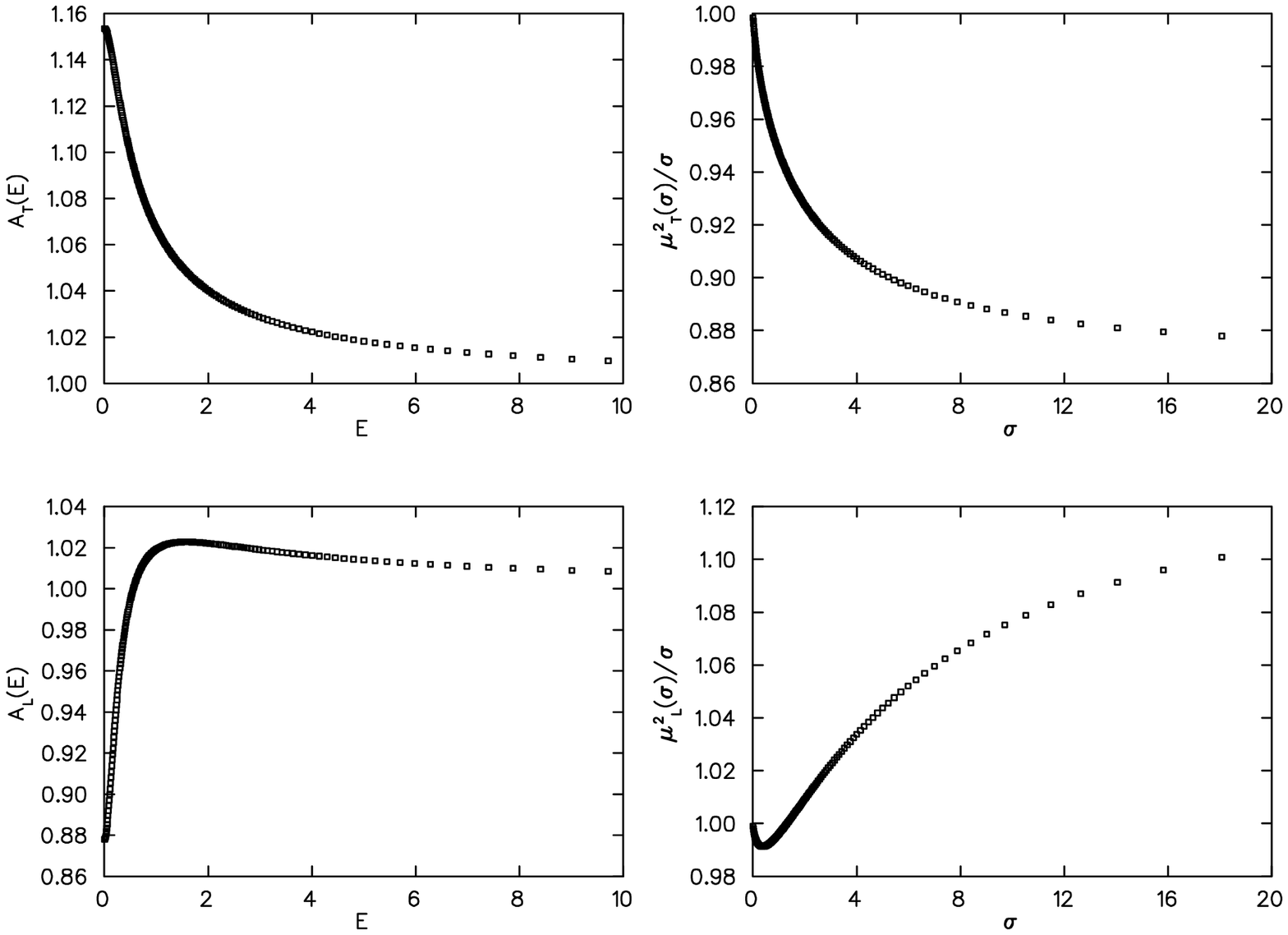}}
\end{center}
\vspace{-0.8cm}
\caption{The profile functions $A_{L,T}(E)$ (left panels)
and pseudotimes $\mu^2_{L,T}$ (right panels) for $\alpha=0.4$}
\label{fig: alpha 0.4 results}
\vspace{0.3cm}
\end{figure}

This compromise is also evident in Fig.~\ref{fig: alpha 0.4 results},
where on the l.h.s. the transverse and longitudinal profile
functions have been plotted as a function of $E$ for $\alpha=0.4$. In
the right hand panels the corresponding pseudotimes, normalized by
their free values, are shown as a function of the proper time
difference $\sigma$.  Note that, curiously, the elongation in the
longitudinal direction does not take place for small values of
$\sigma$.  Rather, in this region of $\sigma$ the dressed nucleon is
`contracted' in all directions as compared to the free case.  This
change in sign of the slope of the (longitudinal) pseudotime (or
equivalently the longitudinal profile functions) is present at all
values of the coupling and, indeed, was also visible in the isotropic
results (see Fig. 2 of Ref.~\cite{WC2}).  This robustness suggests
that this behaviour is not just an artefact of the various trial
actions but, rather, that the variational calculation is attempting to
mimic behaviour present in the exact theory.  In other words, we
suspect that if one were able to calculate the average mean square
displacement in Eq.~(\ref{eq: mean square disp}) with the exact weight
function $e^{i S}$ rather than $e^{iS_t}$ one would find
equivalent behaviour.  This expectation is supported by the fact that
it was precisely this turnover in the profile function which was
required in order to obtain, at least with the isotropic trial actions
discussed in Sec. 4.3 of Ref.~\cite{WC3}, a non-vanishing total
cross section at pion production threshold.

Further evidence that anisotropy is quite important in this problem is
shown in Fig.~\ref{fig: masses}, where the modified bare mass $M_1$ is
plotted as a function of the coupling constant $\alpha$.  Also shown
are the results obtained with the increasingly sophisticated isotropic
trial actions in Ref.~\cite{WC2} (the results have been normalized by
the `worst' of these, namely Feynman's {\it ansatz} for the
retardation function in terms of an exponential, with two variational
parameters).  It is this quantity which serves as a `figure of merit',
because the variational principle guarantees that the true value of
$M_1$ is approached from below.  We see that the improvement in this
bound which is possible with the anisotropic trial action (as compared
to that provided by Feynman's parameterization) is essentially an
order of magnitude larger than that obtained with the best isotropic
action.  Nevertheless, the values of $M_1$ obtained
in this work are at most only about 2 \% above what one obtains with Feynman's
parameterization.  This gives one the hope
that the values of $M_1$ shown in Table~\ref{tab: numerical results}
and plotted in Fig.~\ref{fig: masses} may in fact be quite close to
the exact ones (which, unfortunately, are not known at present but
could be obtained through a direct numerical evaluation of the
functional integral in Eq.~(\ref{eq: propagator})).  We also stress,
because the $M_1$'s obtained here are lower bounds, that this quantity
serves as a useful yardstick to assess and compare the quality of the
different nonperturbative calculations referred to in the
Introduction.

On the other hand, bearing in mind that in ordinary quantum mechanical
applications of the variational principle even small gains in the
ground state energy are usually associated with appreciable
improvements in other observables -- an experience which was also
observed when Feynman's {\it ansatz} for the retardation function was
replaced by the `improved' or `extended' (isotropic) parametrizations
\cite{WC3} -- one might also expect substantial improvement for form
factors and amplitudes when the anisotropic rather than isotropic trial 
actions 
are employed. In particular, some unsatisfactory properties of the
variational description of meson-nucleon scattering, such as the
observed shift in the position of multi-meson thresholds away from 
$s=(M+n m)^2$, the violation of
unitarity \cite{WC5} and the unphysical tails in the scaling function
\cite{WC6} are likely to be amended.

\unitlength1mm
\begin{figure}[ht]
\vspace{-5mm}
\begin{center}
\mbox{\epsfxsize=12cm\epsffile{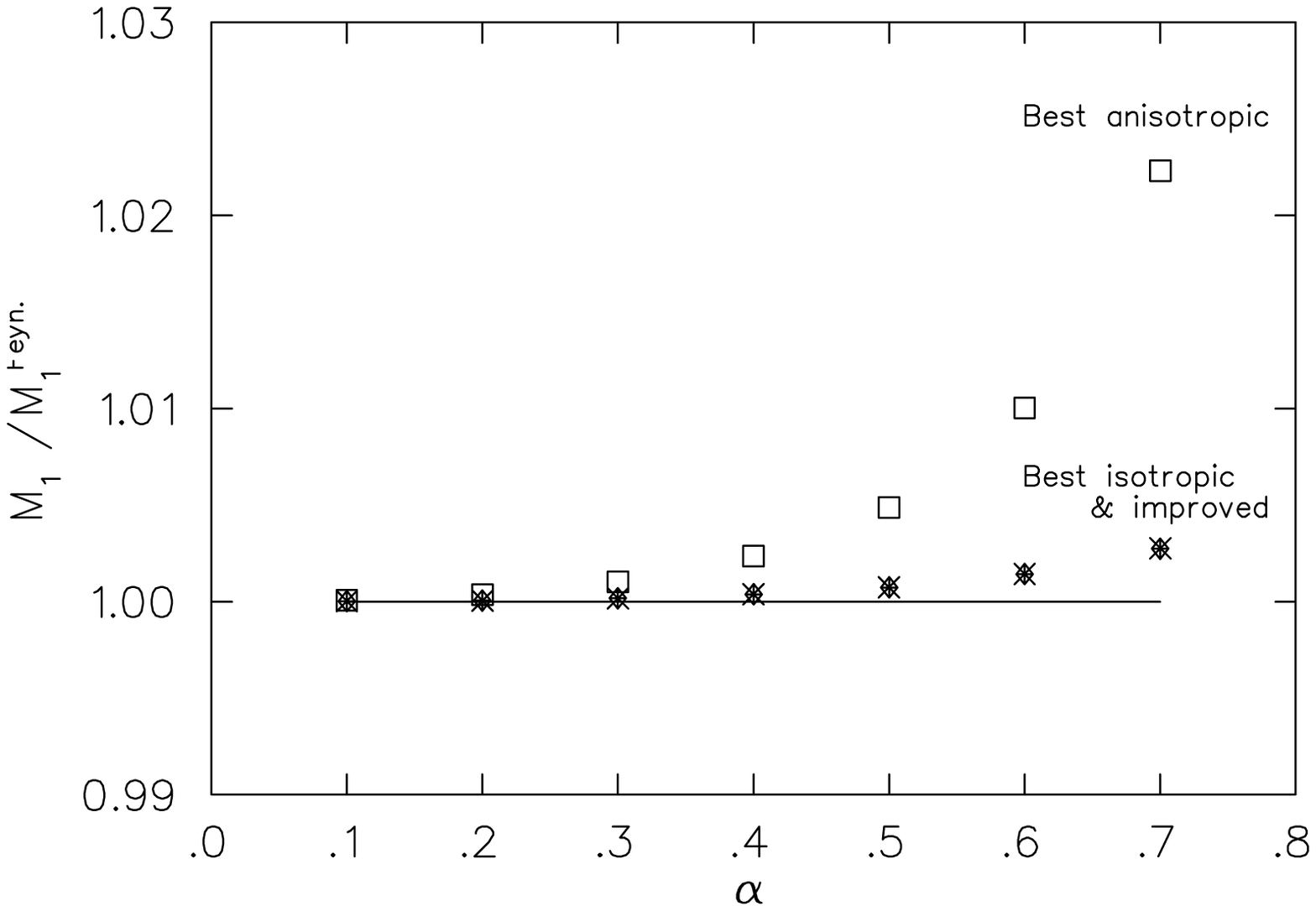}}
\end{center}
\vspace{-0.8cm}
\caption{The finite mass $M_1$ as a function of $\alpha$, normalized by the
result obtained with the Feynman parameterization.  $M_1$ is a strict lower
bound on the exact value.  It is clearly seen that the anisotropic action
leads to significant improvement as compared to the isotropic actions.
These are labelled ``Best'' when no specific form for the retardation function 
has been assumed assuming and ``Improved'' for a parameterization
which incorporates the correct behaviour at small times.}
\vspace{0.3cm}
\label{fig: masses}
\end{figure}

\vspace{0.1cm} 
Finally, we turn to the critical coupling of the
(quenched) theory.  As shown in Ref.~\cite{WC2}, with Feynman's trial
action the variational equations cease to have real solutions above
$\alpha_c \approx 0.824$.  With the best isotropic action this reduces
slightly to $\alpha_c \approx 0.815$.  In the present work, with the
most general anisotropic trial action, we find that this coupling
reduces significantly to $\alpha_c \approx 0.709$. It is remarkable
that if one fixes $\lambda$ to the perturbative value $1$ (i.e. if one
removes it as a variational parameter), the anisotropy in the profile
functions and pseudotimes is now sufficient to produce a critical
coupling, whereas this didn't happen in the isotropic case. Obviously
the occurrence and the magnitude of the critical coupling are directly
linked to the flexibility offered by the trial action and if it would
be possible to go beyond quadratic trial actions one may expect that
the critical coupling would decrease even further.  It is 
tempting to speculate that the true $\alpha_c$ (irrespective of
whether this is $0$ or positive) is being approached from above,
although we know of no rigorous proof of this.  The critical coupling
shows the same (almost) linear rise as a function of the pion mass
which was observed in Ref. \cite{WC2} (see Fig.~\ref{fig: crit coup}),
with the results obtained with the anisotropic action lying about 15\%
below those obtained with the best isotropic action. On the other
hand, the values of the critical coupling observed in the rainbow
Dyson-Schwinger equation studies reported in Ref.~\cite{Alkofer} are
somewhat larger than those labelled 'Best isotropic' in
Fig.~\ref{fig: crit coup}, again with the same linear increase as a
function of $m$.

\unitlength1mm
\begin{figure}[ht]
\vspace{-5mm}
\begin{center}
\mbox{\epsfxsize=12cm\epsffile{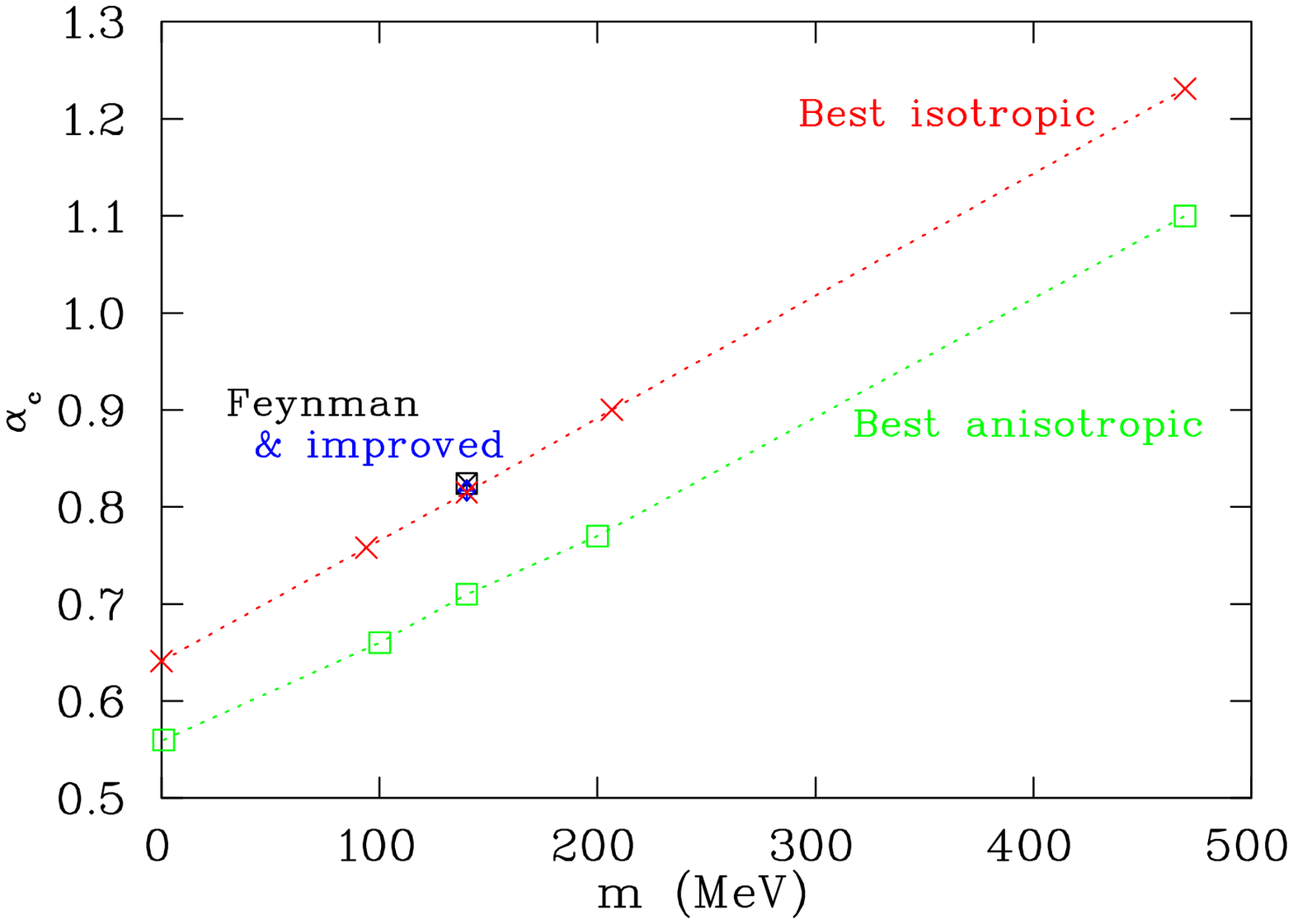}}
\end{center}
\vspace{-0.8cm}
\caption{The critical coupling as function of the pion mass, as obtained
with the anisotropic action.  For comparison, the results of 
Ref.~\protect\cite{WC2} for the isotropic trial actions are also shown 
(the so-called `Feynman' and `improved' actions results were only calculated
at the physical pion mass).  The dotted curves have been drawn to guide the 
eye.}
\label{fig: crit coup}
\vspace{0.3cm}
\end{figure}

\section{Summary and Conclusions}
\label{sec: conclusion}
\setcounter{equation}{0}

We have demonstrated that the most general quadratic trial action in
the polaron variational approach to the scalar Wick-Cutkosky model
leads to a substantial improvement over the previous results which
utilized only isotropic ``profile functions''.  By having a maximum
principle for the intermediate mass $M_1$ , this is clearly reflected
in the values for this quantity displayed in Table~\ref{tab: numerical
results}.  Moreover, the critical value for the coupling constant
$\alpha$ (i.e. the value of $\alpha$ above which no real solutions for
the variational equations exist) obtained with the more general
anisotropic trial action is lower than with the isotropic trial
actions. This is consistent with the interpretation that the more general the
trial action, the closer the critical coupling approaches its exact
value $\alpha_c=0$.
Along the way we have introduced new,
powerful, methods to evaluate various required functional averages, we
have shown that the variational principle itself demands anisotropy
when left enough freedom and discussed the physical picture for the
different evolution of the dressed nucleon parallel or perpendicular
to its momentum.

\vspace{0.1cm} Finally, it may be useful to discuss some distinctions
and merits of the polaron worldline approach compared to other
non-perturbative methods.  Despite its restriction to only quadratic
trial actions this work has again shown that it is capable and
flexible enough to cope with the many different scales occuring in a
field theoretical problem. In addition, we would like to stress its
truly variational aspect which allows to distinguish in a quantitative
manner between the different {\it ans\"atze}; in contrast, the
so-called ``variational'' perturbation theory \cite{varPT} is more an
optimization of perturbation theory with respect to some artificially
introduced parameter and relies on the {\it ad hoc} ``principle of
minimal sensitivity''. Schwinger-Dyson equations are widely used
non-perturbative methods which, however, require truncations of the
infinite hierarchy of equations connecting different Green
functions. This truncation introduces uncertainties which are hard to
control or to quantify and frequently leads to spurious gauge
dependence of results when applied to gauge theories.  In comparison,
the worldline variational approach has been shown to respect gauge
covariance \cite{QED2} and can be corrected systematically by
calculating higher-order cumulants.  Of course, nothing is known about
the convergence of these corrections to the exact result (except that
it seems to be rapid in the polaron case \cite{polaron corr}) but this
feature is shared by any other systematic non-perturbative method.
Note also that no derivative expansion is introduced in the
variational approach as it usually is in the ``Exact Renormalization Group'' 
approach
\cite{ERG} (where it leads to a dependence on the regulator in every
order of that scheme).  Neither has the number of constituents to be
small as in numerical applications of  ``Discretized Light-Cone
Quantization'' \cite{DLCQ} or the Tamm-Dancoff method \cite{TD}.  Again
the polaron provides an instructive example that suggests that such a 
restriction
may be only meaningful for the weak-coupling case: the mean number
of phonons surrounding the dressed electron grows like $ 0.217
\alpha^2 $ at large couplings \cite{Smon}.

Of course, there are also many restrictions of the worldline variational 
method in 
its present form: perhaps the biggest deficiency is its
limitation to the quenched approximation and abelian gauge theories
which prevents application
to realistic strong-coupling theories like Quantum Chromodynamics.
Nevertheless, we believe that this approach has enough virtues to 
make further applications and extensions worthwhile.

\vspace{2cm}

\noindent
{\bf Acknowledgements}: One of us (AWS) would like to thank
Martin Oettel, Will Detmold and Alex Kalloniatis for numerous illuminating
discussions.

\newpage

\bce
{\Large\bf Appendix}
\ece

\renewcommand{\thesection}{\Alph{section}}
\renewcommand{\theequation}{\thesection.\arabic{equation}}

\setcounter{section}{0}

\section{The behaviour of the quenched Wick-Cutkosky model at large orders
of perturbation theory}
\label{app: lopt}
\setcounter{equation}{0}

It is well known that the large order behaviour of a field theory's
perturbative expansion is a useful tool for ascertaining its 
stability~\cite{LOPT reviews}.
In this Appendix we briefly summarize these arguments and apply them
to the quenched cubic scalar theory considered in this paper.

Half a century ago Dyson~\cite{Dyson} argued, on physical grounds,
that the perturbative expansion $\sum f_K\, g^K$ of typical Green
functions, such as the propagator for the field $\Phi$ which we
consider here, should be asymptotic.  Indeed, one typically
finds that the size of the coefficients $f_K$ does grow factorially,
preventing simple-minded summation of the series.  Nevertheless, if
subsequent terms in the expansion oscillate in sign or change in phase
(as they do, for example, in $\Phi^4$ theory~\cite{Lipatov}), Borel
re-summation techniques allow one to restore meaning to the expansion.
Only when all terms at large $K$ contribute with the same sign is this
not possible.  In this case, the perturbative vacuum is not the true
vacuum and tunnelling to the true ground state occurs~\cite{BLZ}.  The
typical, exponentially small, imaginary parts which this entails can
be obtained by first moving the coupling $g$ slightly off the real
axis, therefore allowing the Borel re-summation of the perturbative
expansion, and only at the end analytically continuing back to ${\cal
I}m\, g = 0$.  In this way the divergent nature of an expansion
generates imaginary parts even if $f_K$ is real at all
orders. A good
example of this is provided by \emph {unquenched} $\Phi^3$ theory
which was studied with the techniques outlined here in Refs.~\cite{phi3}.
(Earlier papers discussing the divergent nature of unquenched $\Phi^3$
theory may be found in Ref.~\cite{phi3earlier}.)

Dyson's physical argument involves the different nature of pair production
for positive and negative couplings and clearly is not relevant in a
quenched theory such as the one considered here -- the \emph {vacuum} for
the quenched theory is stable by construction.  This need not be
the case for the particle excitations, however.  Indeed, for the propagator
of $\Phi$, one
can make a rough estimate of $f_K$  by simply counting the number of 
Feynman diagrams at each order, namely 
\be
f_{2K} \> \sim \>\frac{(2 K)!}{2^K K!}\>\sim\> \sqrt 2 \, K^K \, 
\left(\frac{2}{e}\right)^K
\left [ 1 \> + \> O\left(\frac{1}{K}\right)\right]\;\;\;.
\label{eq: rough}
\ee
The typical factorial growth of these coefficients is present in the 
quenched theory just like in the unquenched one.
Moreover, each order contributes with the same sign.  This is readily 
seen from the worldline expression for the $d$-dimensional
Euclidean coordinate space propagator~\cite{WC1}
\bea
G_2(x) &=& {\rm const} \int_0^{\infty} dT \>
\exp \left (- \frac{T}{2} M_0^2 \right )
\> \int_{x(0)=0}^{x(T)=x} {\cal D}x(t) \label{exact 2point(x)}
\nonumber \\
&&\hspace{1cm} \cdot \, \exp \left ( -\int_0^{T} d t  \> \frac{1}{2} \dot x^2 
\> + \>  \frac{g^2}{2} \int_0^{T} dt_1 \>
dt_2 \> < x(t_1) |  \> D_m^{-1} \> | x(t_2) >    \right ) 
\eea
Here $ < x(t_1) |  \> D_m^{-1} \> | x(t_2) >$ is the propagator 
for $\varphi$ and is positive.
In short, unless the $K^K$ behaviour in the above estimate for $f_K$ 
is grossly wrong, the nucleon will unavoidably be
unstable at any coupling, with an exponentially small imaginary 
component to its mass.

The estimate of $f_K$ may be refined through the standard functional
techniques discussed in Refs.~\cite{LOPT reviews} 
(see also Edwards~\cite{edwards} who, by different means, showed 
that the perturbation  expansion of quenched 
$\psi^2 \varphi$ theory is asymptotic  regardless
of whether $\psi$ is fermionic or bosonic).  The method we 
largely follow is that of Itzykson, Parisi and Zuber~\cite{IPZ} who discussed
the vertex function in quenched QED.  In fact, the present application
is much simpler not only because there is no spin and no gauge symmetry
to worry about, but also because in the quenched scalar theory
we need not worry about renormalons or, indeed, UV divergences if
we work in a suitably small spacetime dimension (as pointed out
in Ref.~\cite{Tjon3}, the instability should not be a function of dimension).

The quenched propagator 
\be
G_2(x) \>=\>\frac{\delta}{\delta J(x)}\,\frac{\delta}{\delta J(0)}
\int {\cal D} \varphi {\cal D}\Phi \> \left.
\exp \left \{-\int d^d x \, \left[{\cal L}(x) \,+\,J(x) \Phi(x) \right ]
\right \}_{\rm quenched}\;\;\right|_{J=0}
\label{eq: PI1}
\ee
(${\cal L}$ is given in Eq.~(\ref{def WC}))
is determined by the functional 
average of the propagator of $\Phi$ in a background field generated
by $\varphi$, weighted by the free action associated with $\varphi$:
\be
G_2(x)
 \> = \> \int {\cal D} \varphi \> < x \> | 
\frac{1}{-\Box + M_0^2 - 2g \varphi} |\> y = 0  >
\> \exp \left [ -\frac{1}{2} \int d^d x \,\varphi(x) 
\left (-\Box + m^2\right ) \varphi(x)  \right ] \> \;\;\;,
\label{eq: PI2}
\ee
where we have assumed that the functional integral is appropriately 
normalized.
The largest contribution to the $\varphi$ integral will come from the
region where the denominator in Eq.~(\ref{eq: PI2}) is smallest, i.e.
when 
\be \left [-\Box + M_0^2 - 2g \varphi(x)\right]\,\Phi(x) \>\sim\>
0\;\;\;.  
\ee 
This suggests that we consider the `eigenvalue' equation
\be \left [-\Box + M_0^2 - 2g_n[\varphi] \,\varphi(x)\right ]\,\Phi_n(x)
\>=\> 0\;\;\;, 
\label{eq: diff eq 1}
\ee 
where the index $n$ enumerates the
eigenvalues $g_n$ and eigenfunctions $\Phi_n$, which both have functional
dependence on $\varphi$.  Clearly $g_n$ is antisymmetric, i.e.
\be
g_n[-\varphi]\>=\>-\,g_n[\varphi]\;\;\;
\label{eq: gn odd}
\ee
and will turn out to be nonzero, while the appropriate orthogonality
condition for $\Phi_n$ is
\be
\int d^dx  \> \Phi_i(x)\>  \Phi_j(x)\>\varphi(x) \> = \>  \delta_{ij}\>
\int d^dx \> \Phi^2_i(x)\> \varphi(x)\> \equiv \> 
 \delta_{ij} N_i[\varphi]\;\;\;.
\label{eq: orthogonality}
\ee

We make use of these considerations reverting to the functional
integral  over $\Phi$ in Eq.~(\ref{eq: PI1})  and evaluating it by expanding
 $\Phi$ in terms of the solutions $\Phi_n$, i.e.
\be
\Phi(x) \>=\> \sum_n c_n \Phi_n\;\;\;,
\ee
the functional integral over $\Phi$ being replaced by integrals over
all coefficients $c_n$.
We obtain
\be
G_2(x)
 \> = \> \int {\cal D} \varphi \> \sum_n \frac{\Phi_n(0) \Phi_n(x)}{2 \,
N_n[\varphi] \,(g_n[\varphi]-g)\>}
\> \exp \left [ -\frac{1}{2} \int d^d x \,\varphi(x) 
\left (-\Box + m^2\right ) \varphi(x)  \right ] 
\ee
and so the coefficient of $g^{2K}$ is given by
\be
f_{2 K}
\>=\>   \int {\cal D} \varphi \> \sum_n \frac{\Phi_n(0) \Phi_n(x)}{2 \,
N_n[\varphi] \,g_n[\varphi]}
\> \exp \left [ -\frac{1}{2} \int d^d x \,\varphi(x) 
\left (-\Box + m^2\right ) \varphi(x)  
\> - \> 2 K \log g_n[\varphi]\,\right ] \> \;\;\;.
\label{eq: g2k def}
\ee
Terms odd in $K$ vanish because of the antisymmetry of $g_n$.

For large $K$ the expression for $f_{2 K}$ may be estimated by steepest 
descent in the usual way and hence one seeks solutions $\varphi_0$, leading
to finite action in Eq.~(\ref{eq: g2k def}), of
\be
\left. \left( -\Box\,+\,m^2\right) \varphi \> + \> \frac{2 K}{g_n[\varphi]} \> 
\frac{ \delta}{\delta \varphi}\,g_n[\varphi]\,
\right |_{\varphi=\varphi_0}\> = \> 0\;\;\;.
\ee
The functional derivative of $g_n[\varphi]$ may be evaluated
as in Ref.~\cite{IPZ} and is given by
\be
\frac{ \delta g_n[\varphi]}{\delta \varphi(x)}\>=\>-\,
\frac{g_n[\varphi]\,\Phi^2_n(x)}{N_n[\varphi]}\;\;\;.
\label{eq: diff eq 2}
\ee
The dependence on $K$ may be extracted from the
 coupled set of differential equations~(\ref{eq: diff eq 1}) 
and~(\ref{eq: diff eq 2}) by defining
\be
\varphi_0 \E \sqrt K \, \tilde \varphi_0 \> , \> \> 
\tilde N_n \E \int d^dx \> \Phi^2_n(x)\> \tilde \varphi(x) \> , \> \> 
g_n \E \frac{1}{ \sqrt K} \tilde g_n
\ee
and so the solutions to
\be 
\left (-\Box + M_0^2 \right) \,\Phi_n(x)\,-\, 2 \tilde g_n[\varphi_0] \,
\tilde \varphi_0(x)\,\Phi_n(x) \E  0 
\hspace{1cm}
\left (-\Box + m^2 \right) \,\tilde \varphi_0(x) \> - \> \frac{2 }
{\tilde N_n[\varphi_0]}\, \Phi^2_n(x) \E  0\;\;\;.
\label{eq: diff eq system 2}
\ee
are independent of $K$.  It is not difficult to convince oneself that
normalizable solutions do exist: if one tries 
$\Phi_n = \sqrt{N_n g_n/K}\,\varphi_0$ one obtains
the usual equation for the unquenched theory which has analytic
solutions if $d=6$ and $m=0$~\cite{phi3}.  For $d\ne6$, $m\ne0$ and/or 
general $\Phi$ 
Eq.~(\ref{eq: diff eq system 2}) is readily solved numerically.

One therefore obtains the
large $K$ behaviour of $f_K$ to be essentially the same as that
estimated from diagrammatic counting, i.e.
\bea
f_{2K} \>=\> \sum_n \frac{\Phi_n(0) \Phi_n(x)}{2 \,\tilde N_n[\varphi_0] \,
\tilde g_n[\varphi_0]} \> \frac {K^K}
{\left( e \, \tilde g_n^2[\varphi_0]\right)^K}\> {\rm det}^{-1/2} 
{\cal M}\;\;\;.
\eea
Here $\> {\rm det}\, {\cal M} \> $ is the determinant resulting from the 
quadratic
fluctuations around $\varphi_0$.  Care must be taken in its evaluation 
as the extraction of the zero modes associated with the translational and
dilatation symmetries of Eq.~(\ref{eq: diff eq 2})
introduce extra powers of $\sqrt K$;  we refer the reader to
the extensive literature on this aspect~\cite{LOPT reviews}.  The central 
point, however, is that
the factor $K^K$ already seen in Eq.~(\ref{eq: rough}) remains.

\section{Calculation of Averages}
\label{app: average}
\setcounter{equation}{0}

Here we describe an efficient method to calculate the various averages
needed in the application of the variational principle. It is based on
functional integration over velocities, instead of coordinates, and a
`Hilbert-space' notation for the various proper-time dependent
quantities which allows an easy evaluation of the $T \to \infty$~--~limit
required for the on-mass-shell case. A preliminary account has been
given in Ref. \cite{RAS}.  We will keep the space-time dimension
$d$ arbitrary in these appendices (thereby allowing utilization of
the results for both the 3-dimensional polaron problem as well as dimensionally
regularized theories), while in the main text $d$ has been set to $4$.

\vspace{0.3cm}
\noindent
\subsection{Integration over velocities}
\label{app: velo}

Integration over velocities offers some simplifications, in particular 
for the fermionic case \cite{velo}. In the bosonic case its main
virtue is the absence of explicit boundary conditions and the simple
form the most general quadratic trial action takes if expressed
in velocity variables. Transition to these variables
simply amounts to multiplying the discretized path integral 
\be
\int {\cal D} \tilde x \> e^{i \tilde S[x]} \E {\rm const}
\int d^d (x_b-x_a)  \>  e^{i p \cdot (x_b - x_a) } \, \prod_{k=1}^{N-1} 
\int d^dx_k \> e^{i S[x_k]} \> ,
\hspace{0.2cm} t_b - t_a \E N \Delta t  \> , \> \> x_0 \E x_a \> , 
\> \> x_N \E x_b
\ee
by 
\be
1 = \prod_{k=1}^N \, \int \, d^d v_k \> \delta \left ( \frac{x_k - 
x_{k-1}}{\Delta t} - v_k \right ) \E (\Delta t)^{dN} \prod_{k=1}^N 
\, \int \, d^d v_k \> \delta \left ( x_k - x_{k-1} - \Delta t \, v_k \right ) 
\> . 
\ee
After performing the $x_k$-integrations ($ k = 1, \ldots N-1 $) 
one obtains.
\be
x_k = x_a + \Delta t \sum_{j=1}^k v_j \> , \hspace{1cm}{\rm i.e.} \> \> \> 
x(t) \E x_a + \int_{t_a}^t dt' \> v(t') \> .
\label{x(t) from v(t) 1}
\ee
There is one $\delta$-function left which expresses the constraint 
that full integration over the velocity has to give the final position 
\be
 x(t_b) \E x_b \E x_a + \int_{t_a}^{t_b} dt' \> v(t') \> 
\label{final pos}
\ee
but this is removed by the integration over $x_b - x_a $. Thus
\be
\int {\cal }D \tilde x \> e^{i \tilde S[x]}  \E {\rm const.} \int {\cal }D v 
\> \exp \left ( \> i \tilde S \Bigl [ \, x(t)=x_a+ \int_{t_a}^t dt' \> v(t') 
\, \Bigr ] \> \right ) \> .
\ee
The relation (\ref{x(t) from v(t) 1}) singles out the initial point $x_a$
where the particle starts. Using Eq. (\ref{final pos}) we may also use
the final position $ x_b $ as reference point: 
$ \> x(t) =  x_b - \int_t^{t_b} dt' \, v(t') \> $ .
This can be combined with Eq. (\ref{x(t) from v(t) 1}) into the more 
symmetrical form
\be
x(t) \E \frac{x_b+x_a}{2}  + \frac{1}{2} \, \int_{t_a}^{t_b} dt' \>
{\rm sgn}(t-t') \, v(t')
\label{x(t) from v(t) symm}
\ee
where $ \> {\rm sgn} (x) = 2 \Theta(x) - 1 \> $ is the sign-function.

\subsection{Hilbert space formulation}
\label{app: Hilbert}
It is useful to introduce a kind of Hilbert space notation and 
to write Eq. (\ref{x(t) from v(t) symm}) as
\bea
x(t) \equiv  (t|x) \EA \frac{x_b+x_a}{2}  + \frac{1}{2} (t|S|v) 
\E  \frac{x_b+x_a}{2} + \frac{1}{2} \int_{t_a}^{t_b} dt' \> 
(t|S|t') \, (t'|v) \nonumber \\
\EA  \frac{x_b+x_a}{2} + \frac{1}{2} \int_{t_a}^{t_b} 
dt' \> S(t-t') \, v(t')\> .
\eea
We further define
\be
(t|\circ) \E 1
\label{def circ}
\ee
so that the relation between coordinate and velocity can be
written {\it representation-free} as
\be
|x) \E \frac{x_b + x_a}{2} \,|\circ) + \frac{1}{2} S |v)
\label{x from v}
\ee
Here the sign operator $S$ is defined to have the matrix elements
\be
(t|S|t') \E {\rm sgn} (t-t') 
\ee
and the inner product is defined by integration over the proper time from
$t_a$ to $t_b$. Note that
\be
|\circ) \E S |t = t_a) \E - S |t = t_b)
\label{circ}
\ee
and 
\be 
(t|x) \E  (x|t)^* \E   \frac{x_b + x_a}{2} + \frac{1}{2} (v| S^{\dagger} |t)
\E   \frac{x_b + x_a}{2} + \frac{1}{2}\int_{t_a}^{t_b} dt' \> v(t') \, (-) \, 
{\rm sgn} (t-t')
\ee
since everything is real. Therefore $ S^{\dagger} = - S $
is an antihermitean operator. Since 
$ d \, {\rm sign}(t-t')/dt = 2 \delta (t-t') $
we have, of course, $\dot x(t) = v(t) $. Due to translational invariance we 
are also free to choose the initial and final points: 
$x_a = 0 , x_b = x$ and
$x_a = - x/2 , x_b = x/2$ are most convenient.

\noindent
Let us first consider the previous (isotropic) trial actions in this new 
Hilbert 
space notation: the form (\ref{eq: iso St 1}) can now be written compactly as
\be
S_t \E - \frac{\kappa_0}{2} (v|v) + i \kappa_0^2 
\left ( x \left | f_{\rm diag} - f \right | x \right ) 
\ee
where
\be
\left ( t_1 \left | f_{\rm diag} \right | t_2 \right ) \> := \> 
\delta(t_1 - t_2) \int_{t_0}^{t_0 + T} dt' \> f (t_1 - t') \> .
\ee
Using the relation (\ref{x from v}) this becomes
\be
S_t [v] \E - \frac{\kappa_0}{2} (v| A |v)
\label{S_t}
\ee
where
\be
A \E 1 + \frac{i}{2}  \kappa_0 \, S^{\dagger}  \left ( \, f - f_{\rm diag} 
\, \right ) S \> .
\label{A from f}
\ee
is the ``profile function''.
The same expression for the velocity trial action  is obtained
from Eq. (\ref{eq: iso St 2}), the only difference being that here the 
relation between profile function $A$ and retardation function $g$ 
is
\be
A \E 1 + 2 i \kappa_0 \, g \> .
\label{A from g}
\ee
Actually, Eq. (\ref{S_t}) is also the most general quadratic action 
associated with Eqs. (\ref{eq: aniso St 1}, \ref{eq: aniso St 2}) if 
we understand the profile function as a Lorentz tensor
\be
S_t [v] \E - \frac{\kappa_0}{2} (v_{\mu}| A^{\mu \nu} |v_{\nu}) \> ,
\label{S_t Lorentz}
\ee
i.e. in general the matrix $A$ also has Lorentz indices. 
It is easy to derive its form from  Eqs. 
(\ref{eq: aniso St 1}, \ref{eq: aniso St 2}) but 
we may equally well decompose $A$ from the available structures (the 
metric tensor and the tensor formed from the external momentum $p$), 
forgetting about the original retardation functions.
For later purposes it is most convenient to decompose $A^{\mu \nu}$ 
(and other Lorentz tensors) into components parallel and perpendicular to $p$
\be
A^{\mu \nu}(t - t') \E A_L(t - t') \, \frac{p^{\mu} p^{\nu}}{p^2} 
+ A_T(t - t') \left (  g^{\mu \nu}
- \frac{p^{\mu} p^{\nu}}{p^2} \right ) \> \equiv \>  A_L P_L^{\mu \nu} + 
 A_T P_T^{\mu \nu}  \> .
\label{A par perp}
\ee
If no confusion arises we will not write the Lorentz indices explicitly 
in the following. The extended trial actions (\ref{eq: tilde St}) 
then just add a linear term in $v$ and may be written as
\be
\tilde S_t [v] \E - \frac{\kappa_0}{2} (v| A |v) + ( b | v)
\label{tilde St}
\ee
where
\be
|b ) \E \tilde \lambda \, p \, |\circ) \>.
\label{bt}
\ee
Eq. (\ref{tilde St}), together with Eqs. (\ref{A par perp}, \ref{bt}), 
is the most general linear + quadratic trial action consistent with
translation and time translation invariance. By construction it reduces 
to the free action for $A = \tilde \lambda = 1$.

\subsection{Averages}

\noindent
For evaluation of the various averages needed in Eq. 
(\ref{eq: g2 var}) we use the simple Gaussian 
integral in $d$ dimensions 
\bea
&& \int {\cal D}^d v \> \exp \left \{ \> i \left [ - \frac{\kappa_0}{2} 
(v| A |v) + 
(b | v) \right ] \> \right \} \E  \frac{{\rm const.}}{\det^{d/2} A} \, 
\exp \left [ \> \frac{i}{2 \kappa_0} \left ( b \left | \frac{1}{A} \right |
b \right ) \> \right ] \> =: \> e^{i F}
\nonumber \\
&& F  \E \frac{1}{2 \kappa_0} \left ( b \left | \frac{1}{A} \right | 
b \right )+ \frac{d}{2} i \,  {\rm Tr} \, \log A 
\label{scalar master int}
\eea
as master integral.
Here $ (t|b) $ is an arbitrary function needed to evaluate 
$\langle S_1 \rangle$. A special case is $b(t) = \tilde \lambda p $ needed 
for the average of $ S_t, S_0 $ . The trace implies summation over Lorentz 
indices as well
as integration over the continous proper time.
Eq. (\ref{eq: g2 var}) now becomes
\be
g_2^{\rm var} (p,T) \E \exp \left [ \> i F_t - i F_0 
+ i \left < \,  \tilde S_0 - \tilde S_t + S_1 \, \right > \right ]
\ee
where $F_t, F_0 $ refer to the case of trial and free 
(i.e. $ A = \tilde \lambda  = 1 $) action, respectively.
\phantom{} From the master integral (\ref{scalar master int}) 
we obtain for the 
different averages
\bit
\item[a)] $F_t-F_0$:
\be 
F_t - F_0 \E  \frac{1}{2 \kappa_0} \left [ \, 
\left ( b\left | \left (\frac{1}{A}\right )  \right | b
\right ) - (b_0| b_0 ) \, \right ] 
+ \frac{d}{2} i \, {\rm Tr} \, \log A 
\ee
where $ |b_0) = p | \circ) $.

\item[b)] $ \langle \tilde S_t \rangle $:
\be
\langle \tilde S_t \rangle =   \frac{1}{i} \frac{\partial}{\partial r}
\log \int {\cal D}^d v \, \exp(i r \tilde S_t) \Biggr |_{r = 1}
\E \frac{\partial}{\partial r} F_t [ r A, r b]
\Biggr |_{r = 1} =  \frac{1}{2 \kappa_0} \left ( b \left | 
\frac{1}{A} \right | 
b \right ) +  \frac{d}{2} i {\rm Tr} \, (1)
\ee

\item[c)] $ \langle \tilde S_0 \rangle $: ~~Since 
$ \> \delta A(t_1 - t_2)/\delta A(0) = 
\delta (t_1 - t_2) \> $ we have 
\be
\langle \tilde S_0 \rangle = \left ( \frac{\delta}{\delta A(0)} + 
\int dt \> b_0(t) \frac{\delta}{\delta b(t)}
\right ) F_t = - \frac{1}{2 \kappa_0} \,  
\left ( b \left | \frac{1}{A^2} \right | b \right ) + 
\frac{1}{\kappa_0} \,  
\left ( b \left | \frac{1}{A} \right | b_0 \right ) 
+ \frac{d}{2} i  \, {\rm Tr} \, \frac{1}{A} \> .
\ee

\item[d)]$ \langle S_1 \rangle $: ~~
According to Eq. (\ref{S0+S1}) we have to work out
\be
\Bigl < \> \exp \left \{ \, - i k \cdot \left [ x(t_1) - x(t_2) \right ] \, 
\right \} \> \Bigr > \> .
\ee
The above average may be evaluated again by means of the master integral 
(\ref{scalar master int}) with
\be
b(t) \To b_1(t) \E  \tilde \lambda p  - \frac{k}{2}  \left [ \, 
{\rm sgn}(t_1-t) - {\rm sgn} (t_2-t) \, \right]
\ee
or representation-free
\be
|b_1) \E | b ) + \frac{k}{2} \, S \, \Bigl [ \, 
|t_1) - |t_2) \, \Bigr ] \> .
\label{b_WC}
\ee
We thus obtain
\bea
\left < \> \exp \left \{ \, - i k \cdot \left [ x(t_1) - x(t_2) \right ] \, 
\right \} \> \right > \EA \! \! \exp \left \{ \> \frac{i}{2 \kappa_0} 
\left [ \, 
\left ( b_1 \left | \frac{1}{A} \right | b_1 \right ) - 
\left ( b \left | \frac{1}{A} \right | b \right ) \, \right ] \> \right \}
\nonumber \\
\EA \! \! \exp \left \{ \, \frac{i}{2 \kappa_0} \biggl [\,  
2 a_1 \cdot k + 
k \cdot a_2 \cdot k \, \biggr ] \, \right \} \> =: \> E(k,a_1,a_2)
\label{exp average}
\eea
with
\bea
a_1^{\nu} \EA \frac{1}{2} \Bigl ( b_{\mu} \Bigl | \left ( \frac{1}{A} 
\right )^{\mu \nu} S  \,  \Bigl [ \, \Bigl | t_1 \Bigr)  - \Bigl | t_2 \Bigr )
 \, 
\Bigr ] 
\label{def a1}\\
a_2^{\mu \nu}  \EA \frac{1}{4} \Bigl [ \, \Bigl (t_1 \Bigr | - \Bigl (t_2 
\Bigr | 
\, 
\Bigr ] S^{\dagger}  \left ( \frac{1}{A} \right )^{\mu \nu} S \, \Bigl [ \, 
\Bigl | 
t_1 \Bigr)  - \Bigl | t_2 \Bigr ) \, \Bigr ] \> . 
\label{def a2}
\eea
In $ d = 4 - 2 \epsilon $ dimensions one has to substitute 
$ g^2 \to g^2 \, \nu^{2 \epsilon} $,
where $\nu$ is an arbitrary mass parameter, in order to keep 
$\alpha = g^2/(4 \pi M^2) $ dimensionless in any dimensions. We thus have
\be
\left <S_1 \right>  \E - \frac{g^2}{2 \kappa_0^2} \, \nu^{2 \epsilon}
M^2 \, \int_{t_0}^{t_0+T} dt_1 dt_2 \, \int \frac{d^d k}{(2 \pi)^d} \> 
\frac{1}{k^2-m^2 + i0} \, \exp \left \{ \> \frac{i}{2 \kappa_0} \left [\,  
2 a_1 \cdot k + 
k \cdot a_2 \cdot k \, \right ] \> \right \} \> .
\label{S1 average 1}
\ee
The $k$-integration can be performed by exponentiating the meson 
propagator
\be
\frac{1}{k^2-m^2 + i0} = \frac{1}{2 i \kappa_0} \int_0^{\infty} du' \> 
\exp \left [ \> \frac{i}{2 \kappa_0} \left ( k^2 - m^2 \right ) u' \> 
\right ] \> .
\ee
Taking care of our $\left (+\underbrace{--- \ldots}_{d-1} \right )$-metric in 
Minkowski space we obtain
\bea
\left < S_1 \right>  \EA  - \frac{g^2}{2 \kappa_0^2} \, 
\frac{ \nu^{2 \epsilon}}{2 i \kappa_0} \, 
\left ( - \frac{2 \kappa_0 \pi}{i} \right )^{1/2}
\left ( \frac{ 2\kappa_0 \pi}{i}  \right )^{(d-1)/2} \frac{1}{(2 \pi)^d} 
\,  \int_{t_0}^{t_0+T} dt_1 dt_2
\nonumber \\
&& \hspace{0.2cm} \cdot \int_0^{\infty} du' \> 
\frac{1}{{\rm det}^{1/2} C_{\mu}^{\nu}(u')} \, 
\exp \left \{ \> - \frac{i}{2 \kappa_0} \left [\, m^2 u' +  a_1 
\cdot C^{-1}(u') \cdot a_1  \, \right ] \> \right \} 
\eea
with 
\be
C^{\mu \nu} \E u' g^{\mu \nu} + a_2^{\mu \nu} \E (u' + a_2^L) P_L^{\mu \nu}
+ (u' + a_2^T) P_T^{\mu \nu} \> .
\ee
The decomposition into the orthogonal projectors of
Eq. (\ref{A par perp}) immediately gives
\be
\left ( C^{-1} \right ) ^{\mu \nu}  \E \frac{1}{u'+ a_2^L} P_L^{\mu \nu} +  
\frac{1}{u'+ a_2^T} 
P_T^{\mu \nu}\> , \> \> {\rm det} \, C_{\mu}^{\nu} \E  (u'+ a_2^L) 
\cdot (u'+ a_2^T)^{d-1}  \> .
\ee
Since $a_1  \propto p$ (see Eqs. (\ref{def a1}, \ref{bt})) only the 
longitudinal part of $a_2$ survives in the exponent.
Substituting $ u =  a_2^L/(u'+ a_2^L) $ one then obtains
\bea
\left <S_1 \right>  \EA  \frac{g^2}{16 \pi^2 \kappa_0} \, 
\left ( \frac{ 2 i \pi \nu^2}{\kappa_0} \right )^{\epsilon}  \, 
\int_{t_0}^{t_0+T} dt_1 dt_2 \> \frac{1}{\left ( a_2^L \right)^{1-\epsilon}}
\, \int_0^1 du \> u^{-\epsilon} \nonumber \\
&& \hspace{0.5cm} \cdot \left [ \, 1 + \left ( \frac{a_2^T}{a_2^L} - 1 
\right ) u \, \right ]^{-3/2 + \epsilon} \! \!\cdot 
\exp \left \{ \> - \frac{i}{2 \kappa_0} \left [\, m^2 a_2^L \frac{1-u}{u} +  
\frac{a_1^2}{a_2^L} u \, \right ] \> \right \}  \, .
\label{S1 average 2}
\eea
\eit
\noindent
Putting everything together we have
\bea
\log g_2^{\rm var} (p,T) \! \! \EA  \hspace{-0.3cm} \frac{-i}{2 \kappa_0}
\left [ (b_0 | b_0 )-2 \left ( b \left |\frac{1}{A}\right | b_0 \right )+
\left ( b \left | \frac{1}{A^2} \right| b \right )  \right ] 
  - \frac{1}{2} {\rm Tr} \left [\log A \! +\frac{1}{A}\! -1\right ] 
 + i\left < S_1 \right > 
\label{log g2 var 1}\\
\EA \! - \frac{i T}{2 \kappa_0}
\left \{ p^2 + p \cdot  \left [\, - 2 \tilde \lambda 
\frac{(\circ|A^{-1}|\circ)}{(\circ|\circ)}
+ \tilde \lambda^2 \frac{(\circ|A^{-2}|\circ)}{(\circ|\circ)} \, \right ] 
\cdot p  + 2 \, \Omega (T) + 2\,  V (T) \right \} 
\label{log g2 var 2}
\eea
where
\bea
\Omega(T) \EA   \frac{\kappa_0}{2  i(\circ|\circ)} \, {\rm Tr} 
\left [ \> \log A + \frac{1}{A} - 1 \> \right ] 
\label{Omega(T)}\\
V(T) \EA  - \frac{\kappa_0}{(\circ|\circ)} \left < S_1 \right > \> .
\label{V(T)}
\eea
As indicated, all these quantities are in general $T$-dependent.

\section{Variational equations for finite $T$}
\label{app: finite T}
\setcounter{equation}{0}

Here we present the variational equations for the most general case:
an unspecified profile function 
$A^{\mu \nu}$, a free linear term $b^{\mu}$ in the trial action 
and finite proper time. The latter case is needed, for example, for off-shell
Green functions. Suppressing Lorentz indices and defining
\be
| b ) \E A | c ) \> ,
\label{def c}
\ee
Eq. (\ref{log g2 var 1}) simply becomes
\be
\log g_2^{\rm var} (p,T) \E - \frac{i}{2 \kappa_0}
\Bigl [ (b_0 | b_0 )-2 \left ( c \bigr | b_0 \right )+
\left ( c \bigr | c \right )  \Bigr ] 
  - \frac{1}{2} {\rm Tr} \left [\log A  +\frac{1}{A} -1 \right ] 
 + i \left < S_1 \right >  \> .
\label{log g2 var 3}
\ee
According to Eq. (\ref{def a1}), in the averaged interaction only 
\be
a_1 \E \frac{1}{2} \bigl ( c \bigl | S  \,  
\Bigl [ \, \bigl | t_1 \bigr)  - \bigl | t_2 \bigr ) \, \Bigr ] 
\ee
depends on $|c)$ whereas the sole dependence on $A$
resides in $a_2$ (see Eq. (\ref{def a2})).
Instead of varying with respect to
$b(t)$ or the parameters contained therein we may vary equivalently 
with respect to $c(t)$. As it displays the $a_1, a_2$-dependence in the 
simplest way, the average (\ref{S1 average 1}) is the most convenient 
expression 
for the present purposes (Eq. (\ref{S1 average 2}) is also applicable if 
longitudinal and transverse parts 
are taken with respect to the vector $a_1^{\mu}$ or $c^{\mu}$).
We then obtain the variational equation for $c(t)$ as
\bea
c^{\mu} (t) \EA b_0^{\mu}(t) + \kappa_0
\frac{\delta \left < S_1 \right > }{\delta c_{\mu}(t)} 
\E b_0^{\mu}(t) - \frac{i g^2}{4 \kappa_0^2} \, \nu^{2 \epsilon}
M^2 \, \int_{t_0}^{t_0+T} dt_1 dt_2 \,  \left [ \, {\rm sgn}(t-t_1) - 
{\rm sgn}(t-t_2) \, \right ]  \nonumber \\
&& \hspace{6.5cm} \cdot \, \int \frac{d^d k}{(2 \pi)^d} \> 
\frac{k^{\mu}}{k^2-m^2 + i0} \, E(k,a_1,a_2)  
\label{var eq for c(t)}
\eea
where $E(k,a_1,a_2)$ is defined in Eq. (\ref{exp average}).
Similarly one obtains the variational equation for the profile function by 
varying 
Eq. (\ref{log g2 var 3}) with respect to $ A^{-1}_{\mu \nu}$. This gives
\bea
A^{\mu \nu}(t',t) \EA g^{\mu \nu} \, 
\delta \left (t - t' \right )
- \frac{g^2}{8 \kappa_0^3} \nu^{2 \epsilon} M^2 \, \int_{t_0}^{t_0+T} dt_1 dt_2
\> \left [ \, {\rm sgn}(t-t_1) - {\rm sgn}(t-t_2) \, \right ] \nonumber \\
&& \hspace{1cm} \cdot \left [ \, {\rm sgn}(t'-t_1) - {\rm sgn}(t'-t_2) \, 
\right ] 
\, \int \frac{d^d k}{(2 \pi)^d} \> 
\frac{k^{\mu} k^{\nu}}{k^2-m^2 + i0} \, E(k,a_1,a_2) \> .
\label{var eq for A(t,t')}
\eea
Eqs. (\ref{var eq for c(t)}, \ref{var eq for A(t,t')}) are a system of coupled
nonlinear integral equations which also determine the optimal Lorentz structure
of the linear term and the profile function. Despite the complexity of 
these equations their very structure allows some general observations: first,
it is seen that the profile function is symmetric $ A(t',t) = A(t,t')$. 
Second, we
note that both $c$ and $A$ take their 
free values at the proper time boundaries $t_0, t_0 + T$, since then 
$ {\rm sgn}(t-t_1) - {\rm sgn}(t-t_2) = 0 $.
It is also seen that the inhomogenous term
$b_0^{\mu}(t)$ is essential for the Lorentz structure because it is the only
available 4-vector. If it is absent (for example,
in the case of the massless on-shell propagator) then $c^{\mu} = 0$ and 
$ A^{\mu \nu} \propto g^{\mu \nu}$. This is because 
the $k$-integral in Eq. (\ref{var eq for c(t)})
would be proportional to $a_1^{\mu}$, i.e. $c^{\mu}$ and the 
$k$-integral in Eq. (\ref{var eq for A(t,t')}) 
proportional to $a_1^{\mu} a_1^{\nu}$ and $g^{\mu \nu}$. Therefore, no 
anisotropy
is generated in this case. By the same argument, a constant inhomogenous term
$b_0^{\mu}$ (for example, in the case of the massive on- or off-shell 
propagator
where $b_0^{\mu} = p^{\mu}$) naturally generates 
$c^{\mu}(t) = \lambda(t) b_0^{\mu}$ and
an anisotropic profile function like in Eq. (\ref{A par perp}) where the 
Lorentz decomposition is with respect to the preferred vector  $b_0$. 
In the on-shell
limit $\lambda(t) \to \lambda$ and $ A(t',t) \to A(t-t')$ 
(except at the boundaries) and in $d=4$ Euclidean dimensions we recover 
precisely
the anisotropic variational equations (\ref{eq: var lambda}) - 
(\ref{eq: var AL}).

\section{Large-T limit}
\label{app: large T}
\setcounter{equation}{0}

To obtain the position of the pole in the nucleon propagator
we have to work out the
limit $T \to \infty$ for the various averages calculated above. This is done 
most conveniently by choosing the symmetric proper time interval, i.e.
$t_0 = - T/2 $. 

\subsection{Leading order}
For large $T$ all quantities $ {\cal O}(t - t') $ which only depend on the 
time difference
may be diagonalized in the space of normalized functions
\be
(t|E) \> := \> \frac{1}{\sqrt{2 \pi}} \, e^{ i E t} \> ,
\label{continous basis}
\ee
{\it viz.} 
\bea
(E | {\cal O} | E') \EA \frac{1}{2 \pi} \int_{-T/2}^{+T/2} dt dt' \> 
{\cal O}(t - t') \, \exp ( - i E t + i E' t' ) \nonumber \\
\EA \frac{1}{2 \pi} \int_{-T}^{+T} d\sigma \> 
{\cal O} (\sigma) \, \exp \left ( - i \frac{E+E'}{2} \sigma \right ) \,
\int_{-(T - |\sigma|)/2}^{+(T - |\sigma|)/2} d\Sigma \> e^{- i (E - E') 
\Sigma} 
\label{O_T(E)} \\
&\TO& \delta (E - E' ) \, 
\int_{-\infty}^{+\infty} d \sigma \> {\cal O}(\sigma) \, e^{-i E \sigma} 
\> =: \> \delta (E - E' ) \, \tilde {\cal O}(E) \> .
\label{O(E)}
\eea 
For notational simplicity  we suppress the ``tilde''-sign over the 
Fourier-transformed quantities in the following, i.e. write $A(E)$ for 
$\tilde A(E)$ etc. Using
\bea
(E|\circ) \EA \sqrt{2 \pi} \, \delta(E) 
\label{E circ}\\
S(E) \EA - 2i {\cal P} \frac{1}{E} \> ,
\label{S(E)}
\eea 
where ${\cal P}$ denotes the principal value, 
one can then easily evaluate the required large-$T$ limits. 
For example
\bea
\left ( \circ \left | \left ( \frac{1}{A} \right )^{\mu \nu} \right | 
\circ \right ) \EA 
\int_{-T/2}^{T/2} dt \, dt' \> \frac{1}{2 \pi} \int_{-\infty}^{+\infty} dE \> 
\left ( \frac{1}{A(E)}\right )^{\mu \nu}  \, e^{i E(t-t')} \nonumber \\
\EA \frac{1}{2 \pi} \int_{-\infty}^{+\infty} dE \int_{-T}^T d\sigma 
\int_{-(T - |\sigma|)/2}^{(T - |\sigma|)/2} \! d\Sigma \> 
e^{i E \sigma} \left ( \frac{1}{A(E)}\right )^{\mu \nu}  
\! \TO T   \left ( \frac{1}{A(0)} \right )^{\mu \nu}.
\eea
The last equation can be obtained more easily by using 
Eq. (\ref{E circ})
which gives $ \> 2 \pi \delta(0)/A(0) \> $ and replacing
$ \> 2 \pi \delta(0) \> $ by $T$. This is similar
to calculating a scattering cross section
from the square of a transition matrix which contains an energy-conserving 
$\delta$-function. The same rule applies if one evaluates 
$ ( \circ | \circ ) = T $ 
in the energy representation or $(E|E') = \delta(E-E') \to T/(2 \pi) $ 
for $E \to E'$.

\noindent
Using the traces over Lorentz indices
\be
{\rm tr} \, P_L \E 1  \> \> , \> \> {\rm tr} \, P_T \E  g_{\mu}^{\mu}
- \frac{p_{\mu} p^{\mu}}{p^2} \E d - 1
\ee
one then obtains for the `kinetic term' (\ref{Omega(T)}) 
\bea
\Omega \EA \lim_{T \to \infty} \frac{\kappa_0}{2 i T} \, {\rm tr}  
\int_{-\infty}^{+\infty} dE \> 
(E | E) \, \left [ \, \log A(E) + \frac{1}{A(E)} - 1 \, \right ]
\nonumber \\
\EA  \frac{\kappa_0}{2 i \pi} \, \int_0^{\infty} dE \,
\left \{  \, \log A_L(E) + \frac{1}{A_L(E)} - 1 + (d-1)
\left [ \log A_T(E) + \frac{1}{A_T(E)} - 1 \right ] \, \right \} 
\nonumber \\
&\equiv& \frac{1}{d} \, \Omega [A_L] + \frac{d-1}{d} \,  \Omega [A_T] \, 
\label{Omega aniso}
\eea
where we have used that $A(E) $ is even. This property follows from the 
connection (\ref{A from f}) in $E$-space
\be 
A(E) \, \delta(E-E') \E \delta(E-E') + \frac{i}{2} \kappa_0
\frac{2i}{E} \left ( E \left | f - f_{\rm diag} \right | E' \right )
\frac{-2i}{E'}
\ee
or
\be
A(E) \E 1 + \frac{2 i \kappa_0}{ E^2} \, \int_{-\infty}^{+\infty}
d \sigma \> f(\sigma) \, \left [ \, e^{i E \sigma} - 1 \, \right ]
\E 1 - \frac{8i \kappa_0}{E^2} \, \int_0^{\infty} d \sigma \> f(\sigma)
\sin^2 \left ( \frac{E \sigma}{2} \right ) \> . 
\label{A(E) from f(sigma)}
\ee
Since the retardation function $f(\sigma = t_1 - t_2)$ must be even, it follows
that $A(E) = A(-E) $. The same conclusion is reached if the equivalent form
(\ref{A from g}) is used:
\be
A(E) \E 1 + 4 i \kappa_0 \, \int_0^{\infty} d \sigma \> g(\sigma) \, 
\cos(E \sigma) \> .
\label{A(E) from g(sigma)}
\ee
Note that Eqs. (\ref{A(E) from f(sigma)}, \ref{A(E) from g(sigma)}) are
consistent with the relation $ f(\sigma) = \ddot g(\sigma) $. 

Next we evaluate the coefficients in the exponent of the
averaged interaction term for large $T$ and we obtain by using 
Eqs. (\ref{def a1}, \ref{def a2}, \ref{S(E)})
\bea
a_1^{\nu}(\sigma) &\TO& \frac{1}{2} \int_{-\infty}^{+\infty} dE \> \delta(E) 
\, \tilde \lambda \, p_{\mu} \, \left ( \frac{1}{A(E)} \right )^{\mu \nu} \, 
{\cal P} \frac{-2i}{E} \, \left ( e^{-i E t_1} - e^{-i E t_2} \right )
\nonumber \\
\EA - \sigma \tilde \lambda \, p_{\mu} \left( \frac{1}{A(0)} \right )^{\mu 
\nu} \E  - \sigma \, \frac{\tilde \lambda}{A_L(0)} \,  p^{\nu} \> \equiv \> 
- \sigma \, \lambda p^{\nu}
\label{a1 large T}\\
a_2^{\mu \nu}(\sigma) &\TO& 
\frac{1}{8\pi} \int_{-\infty}^{+\infty} dE \> 
{\cal P} \frac{(-2i)(2i)}{E^2} \, 
\left ( \frac{1}{A(E)}\right )^{\mu \nu} 
\, \left | e^{i E t_1} -e^{i Et_2} \right |^2 \nonumber \\
\EA \frac{4}{\pi} \int_0^{\infty} dE \> 
\frac{\sin^2 E \sigma/2}{E^2} \,  \left ( \frac{1}{A(E)}\right )^{\mu \nu} 
\> \equiv \>  \left ( \mu^2(\sigma) \right )^{\mu \nu} \nonumber \\
\EA \mu^2_L(\sigma) \, P_L^{\mu \nu} + \mu^2_T(\sigma) \, P_T^{\mu \nu} \> .
\label{a2 large T}
\eea
Here we have used the decomposition of $1/A$ into 
orthogonal projectors $P_L , P_T$ and the abbreviation (\ref{eq: def lambda}).
Finally, in the limit $ T \to \infty $ the double integral in $ < S_1 > $ 
simplifies to
\be
\int_{-T/2}^{T/2} dt_1 dt_2 \> \ldots \E \int_{-T}^T d\sigma 
\int_{-(T - |\sigma|)/2}^{+(T - |\sigma|)/2} d\Sigma \> \ldots \TO 
\int_{-\infty}^{+\infty} d\sigma \> \left ( T - |\sigma| \right ) \ldots \E 
2 \int_0^{\infty} d\sigma   \> \left ( T - \sigma \right ) \ldots
\label{double time}
\ee
since the integrand does not depend on $\Sigma = (t_1 + t_2)/2 $ and is even 
in $\sigma$. The potential
term (\ref{V(T)}) therefore becomes
\bea
V &\TO&  - \frac{g^2}{8 \pi^2} \, 
\left ( \frac{ 2 i \pi \nu^2}{\kappa_0} \right )^{\epsilon}  \, 
\int_0^{\infty} d\sigma \> \frac{1}{\left [ \mu^2_L(\sigma) \right]^{1- 
\epsilon}} \, \int_0^1 du \> u^{-\epsilon} \,  
\left [ \, 1 + \left ( \frac{\mu^2_T(\sigma)}{\mu^2_L(\sigma)} - 1 \right ) u 
\, \right ]^{-3/2 + \epsilon} \nonumber \\
&& \hspace{4cm} \cdot 
\exp \left \{ \> - \frac{i}{2 \kappa_0} \left [\, m^2 \mu_L^2(\sigma) 
\frac{1-u}{u} +  \frac{\lambda^2 p^2 \sigma^2}{\mu^2_L(\sigma)} u \, \right ]
 \> \right \}  \> .
\label{V aniso}
\eea
Thus Eq. (\ref{log g2 var 2}) has the following large-$T$ limit
\be
\log \, g_2^{\rm var} (p,T) \TO  - \frac{i}{2 \kappa_0} \left [ \> p^2 
\left ( 1 - \lambda \right )^2 + 2 \, \Omega + 2 \, V \> \right ] \, T 
+ {\cal O} \left ( T^0 \right ) \> .
\label{log g2 large T}
\ee

\subsection{Subasymptotic terms}
\label{app: T NLO}
 
The ${\cal O}(T^0)$-terms are needed for evaluation of the residue of the 
propagator. They can be obtained from 
Eq. (\ref{O_T(E)}) which for finite $T$ reads
\be
(E | {\cal O} | E') \E \frac{1}{2 \pi} \int_{-T}^{+T} d\sigma \> 
{\cal O} (\sigma) \, \exp \left ( - i \frac{E+E'}{2} \sigma \right ) 
\, \frac{\sin [ (E - E') (T - |\sigma|)/2]}{E - E'} \> .
\ee
Since with $y = (T - |\sigma|)/2$
\bea
\int_{-\infty}^{+\infty} dx \> f(x) \, \frac{\sin (x y)  }{x} \EA f(0) 
\int_{-\infty}^{+\infty} dx \> \frac{\sin (x y )}{x} 
+  \int_{-\infty}^{+\infty} dx
\> \frac{f(x) - f(0)}{x} \, \sin (x y ) \nonumber \\
&\TO& \pi f(0) - 2 \frac{f'(0)}{T} + {\cal O} \left ( \frac{1}{T^2} \right )
\eea
one obtains
\be
(E | {\cal O} | E') \E \left [ \, \delta(E-E') - \frac{2}{\pi T} \delta'(E-E') 
+ \ldots \, \right ] \, 
\underbrace{ \int_{-T}^{+T} d\sigma
\>  {\cal O}(\sigma ) \, \exp \left ( - i \frac{E + E'}{2} \sigma \right )}
_{=:{\cal O}_T( (E+E')/2 )} \> .
\ee
As expected, in next-to-leading order the operator ${\cal O}$ is no longer 
diagonal
in $E$-space. More generally, any function of the operator ${\cal O}$ has 
matrix elements 
\be
(E | F({\cal O}) | E') \E  \delta(E-E')  F({\cal O}_T(E)) - 
\frac{2}{\pi T} \delta'(E-E') \>  {\cal O}_T \left ( \frac{E+E'}{2} \right )
 F'\left ( {\cal O}_T \left ( \frac{E+E'}{2} \right )  \right )  + \ldots
\ee
This allows one to evaluate the corrections for large $T$ where we assume that
the integration limits in $ {\cal O}_T$ can be extended to $\pm \infty$ with
impunity (it should be remembered that the retardation functions and 
$A_{L,T}(t-t')$ decay exponentially for
large time differences). However, due to the
$\delta'$-function these corrections involve derivatives of even functions
and therefore lead to a vanishing contribution. For example,
the kinetic term (\ref{Omega(T)})
\bea
\Omega(T) \EA - \frac{\kappa_0}{2 i T} \, {\rm tr} \int_{-\infty}^{+\infty} dE 
dE' \,
\int_{-T/2}^{+T/2} dt \> \left ( E  \left | \log A + \frac{1}{A} -1 \right | 
E' \right ) \, \left ( E' | t \right ) \,  \left ( t | E \right ) \nonumber \\
\EA - \frac{\kappa_0}{2 i T} \, {\rm tr} \int_{-\infty}^{+\infty} dE dE' \> 
\left ( E  \left | \log A + \frac{1}{A} -1 \right | E' 
\right ) \, \frac{1}{\pi} \frac{\sin [ (E-E')T/2 ]}{E-E'}
\eea
does not receive a subasymptotic correction:
\be
\Omega(T) - \Omega  \TO \frac{\kappa_0}{2 i \pi^2 T}  \, {\rm tr} 
\int_{-\infty}^{+\infty} dy \> \left [ \, 1 - \frac{1}{A(y)} \, \right ] \, 
\int_{-\infty}^{+\infty} dx \>  \delta'(x) \, 
\frac{\sin [ x T/2 ]}{x T/2} \,
 \E 0 \> . 
\ee
Similarly the corrections to other quantities either 
lead to odd integrands or to terms involving $A'(0)$. Therefore all 
these corrections vanish and the {\it only} contribution
in next-to-leading order comes from Eq. (\ref{double time})
\bea
V(T) &\TO& V + \frac{1}{T} \,  \frac{g^2}{8 \pi^2} \, 
\left ( \frac{ 2 i \pi \nu^2}{\kappa_0} \right )^{\epsilon}  \, 
\int_0^{\infty} d\sigma \, \frac{\sigma}{\left [ \mu^2_L(\sigma) \right]^{1- 
\epsilon}} \int_0^1 du \, u^{-\epsilon} \,  
\left [ \, 1 + \left ( \frac{\mu^2_T(\sigma)}{\mu^2_L(\sigma)} - 1 \right ) u 
\, \right ]^{-3/2 + \epsilon} \nonumber \\
&& \hspace{4.5cm} \cdot 
\exp \left \{ \> - \frac{i}{2 \kappa_0} \left [\, m^2 \mu_L^2(\sigma) 
\frac{1-u}{u} +  \frac{\lambda^2 p^2 \sigma^2}{\mu^2_L(\sigma)} u \, \right ]
 \> \right \}  \> .
\label{V next}
\eea

\section{Virial theorem}
\label{app: virial}
\setcounter{equation}{0}

Here we prove that the `kinetic term'
\be
\Omega \E \omega \cdot \int_0^{\infty} dE \> \left [ \, \log A(E) + 
\frac{1}{A(E)} - 1 \, \right ]
\ee
may be expressed in terms of the (unspecified) `potential' 
$V [\, \mu^2(\sigma) \, ]$.
 This relation only holds if the variational equation
\be
\delta \left ( \Omega + V \right ) \E 0 \> \> \Rightarrow 
A(E) \E 1 - \frac{1}{\omega} A^2(E) \frac{\delta V}{\delta A(E)}
\label{var eq}
\ee
is fulfilled. The value of the constant $\omega$ will turn out to be 
irrelevant 
and for simplicity we consider here only the isotropic case, the 
generalization to the anisotropic one being straightforward.
 
We first note from the definition (\ref{eq: pseudotimes}) of the pseudotime 
that
\be
\frac{\delta \mu^2(\sigma)}{\delta A(E)} \E - \frac{4}{\pi} 
\frac{\sin^2(E\sigma/2)}{E^2 A^2(E)}
\ee
and therefore by multiplication with $A(E)$
\be
\int_0^{\infty} dE \> A(E) \, \frac{\delta \mu^2(\sigma)}{\delta A(E)} \E 
- \mu^2(\sigma) \> .
\label{help}
\ee
Next we split up $\Omega = \Omega_1 + \Omega_2 $ into two parts and evaluate 
them separately. We write
\be 
\Omega_1 = 
\omega \cdot \int_0^{\infty} dE \> \left [ \frac{1}{A(E)} - 1 \right ] 
\ee
and insert the variational solution (\ref{var eq}) divided by $A(E)$ into 
that expression. In this way one obtains
\bea
\Omega_1^{\rm var} \EA  \int_0^{\infty} dE \> A(E) \, 
\frac{\delta V}{\delta A(E)} \E \int_0^{\infty} dE \> A(E) \,
\int_0^{\infty} d\sigma \> \frac{\delta V}{\delta \mu^2(\sigma)} \, 
\frac{\delta \mu^2(\sigma)}{\delta A(E)} \nonumber \\
\EA - \int_0^{\infty} d\sigma \> \mu^2(\sigma)
\frac{\delta V}{\delta \mu^2(\sigma)} 
\eea
where Eq. (\ref{help}) has been used in the last line. 
After an integration by parts the logarithmic term becomes
\be
\Omega_2 \E - \omega \cdot \int_0^{\infty} dE \> E \frac{A'(E)}{A(E)} \> .
\ee
Here we have assumed for simplicity that the boundary terms do not give a 
contribution but it can be shown that the final outcome is the same even if 
this is not the case. $\Omega_2$ can be
transformed by using the variational equation differentiated with respect 
to the variable $E$
\be
A'(E) \E \frac{4}{\omega \pi} \int_0^{\infty} d\sigma \>
\frac{\delta V}{\delta \mu^2(\sigma)} \, \frac{\partial}{\partial E}
\left ( \frac{\sin^2(E\sigma/2)}{E^2} \right ) \> .
\ee
Converting the derivative with respect to $E$ into one with respect to 
$\sigma$ one obtains
\be
\frac{E A'(E)}{A(E)} \E \frac{1}{\omega} \int_0^{\infty} d\sigma \>
\frac{\delta V}{\delta \mu^2(\sigma)} \, 
\sigma^3 \frac{\partial}{\partial \sigma}
\left ( \frac{1}{\sigma^2} \frac{4}{\pi} \frac{\sin^2(E\sigma/2)}{E^2 A(E)} 
\right )
\ee
and therefore
\be 
\Omega_2^{\rm var} \E - \int_0^{\infty} d\sigma \> 
\frac{\delta V}{\delta \mu^2(\sigma)} \, \sigma^3 \frac{\partial}{\partial 
\sigma}
\left (  \frac{\mu^2(\sigma)}{\sigma^2} \right ) \> .
\ee
Combining both expressions we have
\be
\Omega^{\rm var} \E  - \int_0^{\infty} d\sigma \> 
\frac{\delta V}{\delta \mu^2(\sigma)} \, \sigma^2 \frac{\partial}{\partial 
\sigma}
\left (  \frac{\mu^2(\sigma)}{\sigma} \right ) \E 
\int_0^{\infty} d\sigma \> 
\frac{\delta V}{\delta \mu^2(\sigma)} \, \left [ \, \mu^2(\sigma) - \sigma
\frac{\partial \mu^2(\sigma)}{\partial \sigma} \, \right ] \> .
\label{Omega vir}
\ee
It is instructive to compare this result with the usual virial theorem in 
nonrelativistic quantum mechanics which states that for one particle moving 
in a potential $V(x)$ the expectation value of the 
kinetic energy is given by $ \left < x V'(x) \right > /2 $. 
Since in lowest order $\mu^2(\sigma) = \sigma$ one also sees that the 
variational kinetic energy is proportional to (coupling constant)$^2$ for 
small coupling. Finally, it should be noted that the proportionality constant 
$\omega$ does not appear in Eq. (\ref{Omega vir}). Therefore the result can be 
taken over immediately to the anisotropic trial action and gives 
Eq. (\ref{virial}).


\newpage

\begin{thebibliography}{99}

\bibitem{Feyn} R. P. Feynman, \PRstyle{\PR}{}{97}{1955}{660}.

\bibitem{WC1}  R. Rosenfelder and A. W. Schreiber, 
\PRstyle{\PR}{D}{53}{1996}{3337}.

\bibitem{WC2}  R. Rosenfelder and A. W. Schreiber, 
\PRstyle{\PR}{D}{53}{1996}{3354}.

\bibitem{WC3} A. W. Schreiber, R. Rosenfelder and C. Alexandrou, 
\PRstyle{\IJMP}{E}{5}{1996}{681}.

\bibitem{WC4} A. W. Schreiber and R. Rosenfelder, 
\PRstyle{\NUCPHYS}{A}{601}{1996}{397}.

\bibitem{WC5} C. Alexandrou, R. Rosenfelder and A. W. Schreiber,
\PRstyle{\NUCPHYS}{A}{628}{1998}{427}.

\bibitem{WC6} N. Fettes and R. Rosenfelder, Few-Body Syst. {\bf 24} (1998) 1.

\bibitem{Mano} K. Mano, Progr. Theor. Phys. {\bf 14} (1955) 435.

\bibitem{Wick} G. C. Wick, \PRstyle{\PR}{}{96}{1954}{1124}.

\bibitem{Cut} R. E. Cutkosky, \PRstyle{\PR}{}{96}{1954}{1135}.

\bibitem{Baym} G. Baym, \PRstyle{\PR}{}{117}{1960}{886}.

\bibitem{Alkofer} S. Ahlig and R. Alkofer,
\PRstyle{\ANNPHYS}{}{275}{1999}{113};
R. Alkofer, private communication.

\bibitem{Tjon1} \c{C}. \c{S}avkl{\i}, F. Gross and J. Tjon,
\PRstyle{\PR}{C}{60}{1999}{055210}.

\bibitem{Tjon2} \c{C}. \c{S}avkl{\i}, F. Gross and J. Tjon,
\PRstyle{\PR}{C}{61}{2000}{069901}.

\bibitem{Tjon3} F. Gross, \c{C}. \c{S}avkl{\i} and J. Tjon,
\PRstyle{\PR}{D}{64}{2001}{076008}.

\bibitem{comment} R. Rosenfelder and A. W. Schreiber,
hep-ph/9911484.

\bibitem{MC polaron} C. Alexandrou and R. Rosenfelder, \PRstyle{\PHYSREP}
{}{215}{1992}{1};
J. Titantah {\it et al.}, \PRstyle{\PRL}{}{87}{2001}{206406}.

\bibitem{Bender} C.~M.~Bender, K.~A.~Milton and V.~M.~Savage,
\PRstyle{\PR}{D}{62}{2000}{085001}.

\bibitem{boundstate} For a recent paper on the boundstate problem,
with references to the literature, see
B.~Ding and J.~Darewych, \PRstyle{\JOURPHYS}{G}{26}{2000}{907}; 
for a discussion of the interrelation between the boundstate problem and
the instability, see Ref.~\protect\cite{Alkofer}.

\bibitem{QED2} C. Alexandrou, R. Rosenfelder and A. W. Schreiber,
\PRstyle{\PR}{D}{62}{2000}{085009}.

\bibitem{ARS} C. Alexandrou, R. Rosenfelder and A. W. Schreiber, 
Proceedings of the {\it 15th International IUPAP Conference on Few 
Body Problems in Physics}, Groningen (The Netherlands),
Nucl. Phys. {\bf A631} (1998), 635c .

\bibitem{polaron corr} J. T. Marshall and L. R. Mills, 
\PRstyle{\PR}{B}{2}{1970}{3143}; 
Y. Lu and R. Rosenfelder, 
\PRstyle{\PR}{B}{46}{1992}{5211}.

\bibitem{polaron_aniso} R. Rosenfelder and A. W. Schreiber,
\PRstyle{\PHYSLETT}{A}{284}{2001}{63}.

\bibitem{trunc} G. A. Milekhin and E. S. Fradkin, JETP {\bf 18} (1964) 1323;
B. M. Barbashov {\it et al.}, \PRstyle{\PHYSLETT}{B}{33}{1970}{484};
M. Fabbrichesi {\it et al.},  \PRstyle{\NUCPHYS}{B}{419}{1994}{147}.

\bibitem{QED1} C. Alexandrou, R. Rosenfelder and A. W. Schreiber, 
\PRstyle{\PR}{A}{59}{1999}{1762}.

\bibitem{varPT} see, for example: 
P. M. Stevenson, \PRstyle{\PR}{D}{23}{1981}{2916};
C. M. Bender {\it et al.} \PRstyle{\PR}{D}{45}{1992}{1248};
D. Gromes, \PRstyle{\ZEITPHYS}{C}{71}{1996}{347};
H. Kleinert, \PRstyle{\PR}{D}{57}{1998}{2264};
S. Chiku, Progr. Theor. Phys. {\bf 104} (2000) 1129.

\bibitem{ERG} see, for example: 
C. Bagnuls and C. Bervillier, \PRstyle{\PHYSREP}{}{348}{2001}{91};
J. Polonyi, hep-th/0110026;
D. F. Litim, J. High Energy Phys. {\bf 11} (2001) 059

\bibitem{DLCQ} S. J. Brodsky, J. R. Hiller and G. McCartor,
\PRstyle{\PR}{D}{60}{1999}{054506};
\PRstyle{\PR}{D}{64}{2001}{114023} .

\bibitem{TD} J.R. Spence and J.P. Vary, \PRstyle{\PR}{C}{52}{1995}{1668}. 

\bibitem{Smon} M. A. Smondyrev, Theor. Math. Phys. {\bf 68} (1987) 653.

\bibitem{LOPT reviews} J. Zinn-Justin, Lecture Notes in Physics (Springer)
{\bf 77} (1978) 126; 
\PRstyle{\PHYSREP}{}{49}{1979}{205};
\PRstyle{\PHYSREP}{}{70}{1981}{109}; 
E. Br\'ezin, Proceedings of the 
{\it 1977 European Conference On Particle Physics}, Budapest (Hungary),
eds. L.~Jenik and I.~Montvay, p. 1231;  see also the collection
of papers in {\it Large-Order Behaviour of Perturbation Theory}, eds.
J.~C.~Le Guillou and J.~Zinn-Justin, North Holland (1990).

\bibitem{Dyson}F.~J.~Dyson,\PRstyle{\PR}{}{85}{1952}{631}.

\bibitem{Lipatov}L.~N.~Lipatov, \PRstyle{JETP Lett.}{}{25}{1977}{104}.

\bibitem{BLZ} E. Br\'ezin, J. C. Le Guillou and J. Zinn-Justin,
\PRstyle{\PR}{D}{15}{1977}{1558}.

\bibitem{phi3} A.~Houghton, J.~S.~Reeve and D.~J.~Wallace,
\PRstyle{\PR}{B}{17}{1978}{2956}; 
A.~J.~McKane,\PRstyle{\NUCPHYS}{B}{152}{1979}{166};
\PRstyle{\JOURPHYS}{A}{19}{1986}{453}; 
S.~A.~Newlove,\PRstyle{\JOURPHYS}{A}{17}{1984}{1843}.

\bibitem{phi3earlier}C.~A.~Hurst,
\PRstyle{Camb. Phil. Soc.}{}{48}{1952}{625};
W.~Thirring, \PRstyle{Helv. Phys. Acta}{}{26}{1953}{33};
R.~Utiyama and T.~Imamura, \PRstyle{Prog. Theor. Phys.}{}{9}{1953}{431};
A.~Petermann, \PRstyle{Arch. Sci. Soc. Phys. Hist. Nat. Geneve}
{}{6}{1953}{5}; \PRstyle{Helv. Phys. Acta}{}{26}{1953}{291}.

\bibitem{edwards} S.~F.~Edwards, \PRstyle{Phil. Mag.}{}{45}{1954}{758};
\PRstyle{\it ibid.} {}{46}{1955}{569}.

\bibitem{IPZ} C.~Itzykson, G.~Parisi and J.-B.~Zuber,
\PRstyle{\PR}{D}{16}{1977}{996}.

\bibitem{RAS} R. Rosenfelder, C. Alexandrou and A. W. Schreiber, in: 
Proceedings of the {\it Workshop on Nonperturbative Methods in Field Theory}, 
Adelaide (Australia), eds. A. W. Schreiber, A. G. Williams and A. W. Thomas, 
World Scientific (1998), p. 163.

\bibitem{velo} D. M. Gitman and Sh. M. Shvartsman, hep-th/9310074;
\PRstyle{\PHYSLETT}{B}{318}{1993}{122}; erratum: {\it ibid.} {\bf B331} 
(1994) 449; 
W. da Cruz, \PRstyle{\JOURPHYS}{A}{30}{1997}{5225}.


\end{thebibliography}
\end{document}